\documentclass[a4paper,11pt]{article}
\pdfoutput=1
\usepackage{jheppub}
\usepackage{xcolor,bm,comment,braket}
\usepackage[T1]{fontenc}
\usepackage{comment}
\newcommand{\del}{\partial}
\newcommand{\beq}{\begin{eqnarray}}
\newcommand{\eeq}{\end{eqnarray}}

\newcommand{\tr}{\mathop{\mathrm{tr}}}
\newcommand{\SU}{\text{SU}}
\newcommand{\U}{\text{U}}
\newcommand{\rmi}{i}
\newcommand{\rme}{e}
\newcommand{\rmd}{d}

\newcommand{\bs}{\boldsymbol}
\newcommand{\bb}{\mathbb}
\newcommand{\fr}{\frac}

\newcommand{\der}{\partial} 

\newcommand{\mtx}[1]{\left(\begin{matrix} #1 \end{matrix}\right)}

\newcommand{\sr}{\sqrt}

\begin{document}

\title{
Spin Statistics and Surgeries of Topological Solitons in QCD Matter in Magnetic Field  
}

\author[a]{Yuki Amari}
\emailAdd{amari.yuki@keio.jp}

\author[a,b]{Muneto Nitta}
\emailAdd{nitta@phys-h.keio.ac.jp}

\author[a]{and Ryo Yokokura}
\emailAdd{ryokokur@keio.jp}

\affiliation[a]{Department of Physics \& 
Research and Education Center for Natural Sciences,
Keio University, 4-1-1 Hiyoshi, Kanagawa 223-8521, Japan}
\affiliation[b]{
International Institute for Sustainability with Knotted Chiral Meta Matter(SKCM$^2$), Hiroshima University, 1-3-2 Kagamiyama, Higashi-Hiroshima, Hiroshima 739-8511, Japan
}

\abstract{
The ground state of QCD with two flavors (up and down quarks) at finite baryon density in sufficiently strong magnetic field is in a form of either a chiral soliton lattice(CSL), an array of solitons stacked along the magnetic field, or 
a domain-wall Skyrmion phase in which 
Skyrmions are spontaneously created on top of the CSL. 
In the latter, one 2D (baby) Skyrmion in the chiral soliton corresponds to two 3D Skyrmions (baryons) in the bulk.  
In this paper, we study spin statistics of 
topological solitons 
by using the following two methods: 
the conventional Witten's method 
by embedding the pion fields of two flavors into those of 
three flavors with the Wess-Zumino-Witten (WZW) term,  
and a more direct method by using 
the two-flavor WZW term 
written in terms of a spin structure.
We find that 
a chiral soliton of finite quantized size called 
a pancake soliton 
and a hole on a chiral soliton are fermions or bosons 
depending on 
odd or even quantizations of  
their surface areas, respectively, 
and a domain-wall Skyrmion is a boson.
We also propose surgeries of topological solitons: 
a domain-wall Skyrmion (boson) can be cut into 
a pancake soliton (fermion) and a hole (fermion), 
and 
a chiral soliton without Skyrmions 
can be cut into 
a pancake soliton (fermion) and a hole (fermion).
}

\maketitle

\section{Introduction}

Spin statistics says that a wave function 
or a Hilbert space 
is invariant under exchanges of two bosons 
while it acquires a minus sign under exchange of two fermions. 
Bosons and fermions obey Bose-Einstein and Fermi-Dirac statistics, respectively.
Anyons can exist in lower dimensions. 
 In the case of elementary particles or their composites, 
 their statistics are a priori given 
 by their definition.
However, once one considers solitons, the problem becomes quite subtle.
For instance, 
magnetic monopoles are fermions \cite{Shnir:2005vvi}, and 
Skyrmions \cite{Skyrme:1962vh} 
are fermions or bosons for odd or even number $N_{\rm C}$ of an $\SU(N_{\rm C})$ color gauge group, respectively  \cite{Finkelstein1968,Witten:1983tx} (see also refs.~\cite{ZAHED19861,makhankovskyrme,Manton_Skyrmionbook} as a review). 
In an O(3) model in 2+1 dimensions, 
topological lumps are anyons in the presence of the Chern-Simons (CS)-like term \cite{Wilczek:1983cy,Wu:1984kd}.
 
Here we are interested in particular in Skyrmions. 
When chiral symmetry undergoes spontaneous breaking
in Quantum Chromodynamics (QCD), 
the emergence of massless Nambu-Goldstone (NG) bosons or pions, which are dominant at low energy, is observed. This low-energy dynamics can be described by either the chiral Lagrangian or the chiral perturbation theory (ChPT) centered on the pionic degree of freedom. Importantly, this description is predominantly dictated by symmetries and only modulated by certain constants, including the pion's decay constant $f_{\pi}$ and quark masses $m_{\pi}$  \cite{Scherer:2012xha,Bogner:2009bt}.
Skyrmions are topological solitons 
supported by the third homotopy group 
$\pi_3 [{\rm SU}(N_{\rm F}) ] \simeq  {\mathbb Z}$ 
in the chiral Lagrangian \cite{Skyrme:1962vh} 
where $N_{\rm F}$ is a number of flavors,
and they are 
proposed to describe baryons. 
In QCD, baryons are 
composite particles consisting of 
$N_{\rm C}$ quarks 
in ${\rm SU}(N_{\rm C})$ QCD, 
and thus they are 
fermions or bosons 
for odd or even  $N_{\rm C}$, respectively. 
In order for Skyrmions to be identified with 
baryons, they have to exhibit such spin statistics. 
Witten clarified this problem 
in his elegant paper \cite{Witten:1983tx},
for $N_{\rm F}$ flavors of quarks 
with $N_{\rm F} \geq 3$: 
the Wess-Zumino-Witten (WZW) term 
topologically supported by $\pi_5 [{\rm SU}(N_{\rm F})] \simeq
{\mathbb Z}$ for $N_{\rm F} \geq 3$ 
is present in a response to the chiral anomaly in QCD.
This term gives a correct sign $(-1)^{N_{\rm C}}$ 
in front of the action 
in the presence of a Skyrmion adiabatically rotated 
by $2\pi$
so that 
the Skyrmion is shown to be a fermion or boson 
for odd or even $N_{\rm C}$, respectively.  
This elegant description breaks down 
for two flavors $N_{\rm F} = 2$ because
of the absence of the WZW term 
with $N_{\rm F} = 2$ due to 
$\pi_5 [{\rm SU}(N_{\rm F}=2)] =0$. 
In order to overcome this problem for 
$N_{\rm F} = 2$, Witten proposed 
to embed 
the ${\rm SU}(2)$ chiral Lagrangian to ${\rm SU}(3)$ one 
to obtain the correct statistics of a Skyrmion 
\cite{Witten:1983tx} (see also \cite{Elitzur:1984kr}).

Recently
the WZW term for $N_{\rm F} = 2$ 
associated with a spin structure
 \cite{Lee:2020ojw}
has been found. 
 Although this $N_{\rm F} = 2$ WZW term is perturbatively absent in a flat Minkowski spacetime,
 it gives a nontrivial contribution when one considers spin statistics:
one compactifies the time direction to $S^1$ 
and adiabatically moves a Skyrmion by $2\pi$ in that direction. 
By doing so, one finds that the WZW term gives the correct sign $(-1)^{N_{\rm C}}$ in front of the action:
it is a fermion or boson for even odd or even $N_{\rm C}$, respectively. 

The purpose of this paper is to clarify spin statistics of Skyrmions and chiral solitons in QCD with two flavors (up and down quarks) at finite density 
in the presence of strong magnetic field.
Recently, the QCD phase diagram, especially under extreme conditions like high baryon density, pronounced magnetic fields, and rapid rotation, garners significant attention \cite{Fukushima:2010bq}. 
As an extreme condition,
QCD in strong magnetic fields has received quite intense attention because of the interior of neutron stars
and heavy-ion collisions.
In the presence of an external magnetic field $B$
at finite chemical potential $\mu_{\textrm{B}}$,
the chiral Lagrangian is accompanied by 
the WZW term containing  
an anomalous coupling of the neutral pion $\pi^0$ to the magnetic field via the chiral anomaly 
\cite{Son:2004tq,Son:2007ny}
in terms of 
the Goldstone-Wilczek current \cite{Goldstone:1981kk,Witten:1983tw}.
It was determined to reproduce the so-called chiral separation effect  \cite{Vilenkin:1980fu,Son:2004tq,Metlitski:2005pr,Fukushima:2010bq,Landsteiner:2016led} in terms of the neutral pion $\pi^0$.
Then, under a sufficiently strong magnetic field $B$ 
satisfying 
$B  
    \geq B_{\rm CSL} =
    {16\pi m_{\pi}/f_{\pi}^2}{e \mu_{\textrm{B}}}$, 
the ground state of QCD with two flavors  becomes inhomogeneous  
in the form of a chiral soliton lattice (CSL) 
consisting of a stack of 
domain walls or solitons 
carrying a baryon number 
\cite{Son:2007ny,Eto:2012qd,Brauner:2016pko}.
While these results are based on 
zero temperature analyses, 
it was also shown that 
thermal fluctuations enhance their 
stability~\cite{Brauner:2017uiu,Brauner:2017mui,Brauner:2021sci,Brauner:2023ort}.\footnote{ 
In the context of QCD, 
similar CSLs composed on 
the $\eta$ or $\eta'$ meson 
also appear 
in the case under rapid rotation~\cite{Huang:2017pqe,Nishimura:2020odq,Chen:2021aiq,Eto:2021gyy,Eto:2023tuu,Eto:2023rzd}.
Further investigations have been done 
into the quantum nucleation of CSLs in~\cite{Eto:2022lhu,Higaki:2022gnw},  
quasicrystals \cite{Qiu:2023guy}, 
and the interplay between Skyrmion crystals at zero magnetic field and the CSL ~\cite{Kawaguchi:2018fpi,Chen:2021vou,Chen:2023jbq}.
}
However,
such a CSL state has two kinds of instabilities, and is replaced by another state.
One is 
a charged pion condensation (CPC) 
in a region of higher density and/or stronger magnetic field, 
asymptotically expressed 
at large $B$ as $B \geq B_{\rm CPC} 
    \sim {16 \pi^4 f_{\pi}^4}/{\mu_{\rm B}^2}$ \cite{Brauner:2016pko}  
above which tachyon appears and the CSL becomes  unstable.
 An Abrikosov's vortex lattice was proposed as a consequence of the CPC~\cite{Evans:2022hwr,Evans:2023hms}.
More relevant instability of the CSL occurring 
below the CPC instability  is 
a transition to 
a domain-wall Skyrmion (DWSk) phase in a region 
$B \geq B_{\rm DWSk}(< B_{\rm CPC})$
in which Skyrmions are created on top of the solitons 
in the ground state \cite{Eto:2023lyo,Eto:2023wul}.
These two instability curves meet 
at a single tricritical point
    $(\mu_{\rm c},B_{\rm c}) 
    = 
    \left({16\pi f_{\pi}^2}/{3m_{\pi}}, 
    {3m_{\pi}^2}/{e} \right) 
    \approx 
    \left(1.03 \;\; {\rm GeV}, 
    0.06 {\rm GeV}^2\sim 1.0\times 10^{19} {\rm G}\right)$ 
    on the critical curve 
    $ B= B_{\rm CSL}$ of the CSL phase.
Domain-wall Skyrmions are composite states of a domain wall and Skyrmions, initially introduced in the  field theoretical models in 3+1 dimensions~\cite{Nitta:2012wi,Nitta:2012rq,Gudnason:2014nba,Gudnason:2014hsa,Eto:2015uqa,Nitta:2022ahj} and in 2+1 dimensions~\cite{Nitta:2012xq,Kobayashi:2013ju,Jennings:2013aea}.\footnote{
The term ``domain-wall Skyrmions'' can be traced back to its initial use in ref.~\cite{Eto:2005cc}, where it describes Yang-Mills instantons absorbed into a domain wall in 4+1 dimensions. 
} In condensed matter physics, a 2+1 dimensional variant has been theoretically investigated~\cite{PhysRevB.99.184412,KBRBSK,Ross:2022vsa,Amari:2023gqv,Amari:2023bmx,PhysRevB.102.094402,Kim:2017lsi,Lee:2022rxi} and experimentally observed in chiral magnets~\cite{Nagase:2020imn,Yang:2021}, 
in which case a baby Skyrmion 
supported by $\pi_2(S^2) \simeq {\mathbb Z}$
becomes a sine-Gordon soliton supported by 
$\pi_1(S^1) \simeq {\mathbb Z}$
in the domain-wall effective theory 
which is a sine-Gordon model.
By contrast, in QCD 
in 3+1 dimensions, 
Skyrmions supported by $\pi_3(S^3) \simeq {\mathbb Z}$
are absorbed into a chiral soliton 
to become 
topological lumps (or baby Skyrmions) 
supported by $\pi_2(S^2)\simeq {\mathbb Z}$ in 
an O(3) sigma model 
or the ${\mathbb C}P^1$ model,
constructed by the moduli approximation \cite{Manton:1981mp,Eto:2006uw,Eto:2006pg}
as 
the effective worldvolume theory on a soliton \cite{Eto:2023lyo}. 
One of the important features is that one lump 
 in the soliton 
 corresponds to two Skyrmions in the bulk.
Domain-wall Skyrmions in multiple chiral solitons 
(that is a CSL) 
are Skyrmion chains \cite{Eto:2023wul}, 
giving the more precise phase boundary between 
the CSL and DWSk phases.

In this paper, we investigate spin statistics of 
topological solitons 
in QCD with $N_{\rm F}=2$ 
at finite baryon density $\mu_{\rm B}$
in magnetic field $B$:
a domain-wall Skyrmion, 
a chiral soliton of a finite size called a pancake soliton, 
and a hole of a chiral soliton. 
The pancake soliton has 
a quantized surface area that is the integer multiple of 
the minimally quantized surface area $S_0=2\pi/e B$ \cite{Son:2007ny}. 
Since the conventional WZW term vanishes for the two flavors, 
we follow the two methods;
 one is the Witten's method 
 to embed the $\SU(2)$ chiral field 
 to the $\SU(3)$ one \cite{Witten:1983tx,Elitzur:1984kr} and 
 the other is to use the $N_{\rm F}=2$ WZW term 
written in terms of a spin structure
 \cite{Lee:2020ojw}. 
Although the $N_{\rm F}=2$ WZW term vanishes in flat Minkowski spaces, 
 it can determine spin statistics of topological solitons.  
We find that 
the pancake solitons and holes are fermions (bosons) 
if their area are odd (even) multiple of $S_0$,
while domain-wall Skyrmions 
are bosons.
The latter can be shown alternatively 
by constructing  
a term induced from the $N_{\rm F} = 2$ WZW term
in the domain-wall effective theory. 
We also propose surgeries of topological solitons:
a domain-wall Skyrmion which is a boson can be decomposed into 
a pancake soliton and a hole, 
both of which are fermions. 
Likewise 
a chiral soliton without Skyrmions 
can be decomposed into 
a pancake soliton and a hole, 
both of which are fermions.

This paper is organized as follows.
In sec.~\ref{sec:chiral-Lagrangian} 
we briefly summarize 
the chiral Lagrangian of two flavors 
with the WZW term 
under strong magnetic fields.
In sec.~\ref{sec:topo-solitons}
we introduce topological solitons 
in strong magnetic fields: 
CSL, 
domain-wall Skyrmions, 
neutral pion strings, 
pancake solitons, 
and holes on a chiral soliton. 
We also propose surgeries of topological solitons 
in Sec.~\ref{sec:surgery}.
In sec.~\ref{sec:spin-statistics},
we 
discuss spin statistics of  
pancake solitons, 
holes on a chiral soliton,
and 
domain-wall Skyrmions 
by using 
the conventional Witten's method 
as well as the $N_{\rm F}=2$ WZW term formulated by 
the spin structure. 
As for domain-wall Skyrmions, 
alternatively we also use the effective theory on a single chiral soliton with the term induced from 
the $N_{\rm F}=2$ WZW term.
Section \ref{sec:summary} is devoted to a summary and discussion.
In Appendix \ref{WZW5} we give some details of the evaluation of the WZW term 
in the background of topological solitons.

\section{Chiral Lagrangian}\label{sec:chiral-Lagrangian}
In this section, we summarize the chiral Lagrangian or ChPT at the leading order
${\cal O}(p^2)$
in order to fix notations.
We focus on the phase where chiral symmetry is spontaneously broken.
The effective field theory of pions, known as ChPT, can describe the low-energy dynamics.
The pion fields are represented by a $2\times 2$ unitary matrix,
\begin{equation}
    \Sigma = \rme^{\rmi \tau_a \chi_a} \,,
\end{equation}
where $\tau_a$ (with \(a = 1, 2, 3\)) are the Pauli matrices,
normalized as \(\tr(\tau_a \tau_b) = 2 \delta_{ab}\).
The field \(\Sigma\) transforms under the \(\SU(2)_{\mathrm{L}} \times \SU(2)_{\mathrm{R}}\) chiral symmetry as
\begin{equation}
    \Sigma \rightarrow L \Sigma R^{\dagger} \,,
\end{equation}
where both \(L\) and \(R\) are \(2 \times 2\) unitary matrices. 
The field \(\Sigma\) can be parameterized by
\begin{eqnarray}
&&\Sigma
 = 
\mtx{
\sigma + i\pi^0 & i(\pi^1 - i \pi^2)  
\\ 
i(\pi^1 + i \pi^2)  & \sigma - i\pi^0   
}
 = 
\mtx{
\sigma + i\pi^0 & i\pi^+  
\\ 
i\pi^-  & \sigma - i\pi^0   
},
\end{eqnarray}
 in terms of the charged pions 
 $\pi^\pm \equiv \pi^1 \mp i \pi^2$ 
 and the neutral pion  $\pi^0$.
The \( \U(1)_{\mathrm{EM}} \) 
electromagnetic gauge 
transformation is given by
\begin{eqnarray}
&&    \Sigma \to \rme^{iQ\lambda} \Sigma \rme^{-iQ\lambda} = \rme^{i\lambda\frac{\tau_3}{2}} \Sigma \rme^{-i\lambda\frac{\tau_3}{2}} ,
  \quad \pi^\pm \to e^{\pm i\lambda}\pi^\pm\,,
    \\ 
&& \quad A_\mu \to A_\mu - \frac{1}{e} \partial_\mu \lambda 
\end{eqnarray}
with the electric charge matrix $Q$ of quarks given by
\begin{equation}
    Q = \frac{1}{6}\bm{1}_2 + \frac{1}{2}\tau_3.
\end{equation}
The covariant derivative $D_{\mu}$ 
associated with this transformation 
is given by
\begin{gather}
    D_{\mu}\Sigma
    \equiv \del_{\mu}\Sigma
    + \rmi e A_{\mu} [Q, \Sigma]
    = \del_{\mu}\Sigma
    + \frac{\rmi e}{2} A_{\mu} [\tau_3,\Sigma]
    \label{def_cov_del} \,, 
    \quad 
D_\mu\pi^\pm = (\der_\mu \pm \rmi e A_\mu) \pi^\pm   .
\end{gather}

Our notations in this paper are
 $\eta_{\mu\nu} = (+,-,-,-)$ and 
 $\epsilon^{0123} = \epsilon^{123} = - \epsilon_{123} = 1$.
Then, the effective Lagrangian at the ${\cal O}(p^2)$ order of ChPT can be written as 
\begin{equation}
    \mathcal{L} = \mathcal{L}_{\mathrm{ChPT}} + \mathcal{L}_{\mathrm{WZW}} 
    \label{eq:total-L}
\end{equation}
where the first term is ($\mu=0,\cdots,3$)
\begin{gather}
    \mathcal{L}_{\textrm{ChPT}}
    = \frac{f_{\pi}^2}{4} \tr \left(D_{\mu}\Sigma D^{\mu}\Sigma^{\dag} \right) 
    - \frac{f_{\pi}^2m_{\pi}^2}{4}
    \tr 
    (2 {\bf 1}_2 -\Sigma-\Sigma^{\dag}) \label{ChPT_with_B} \,,
\end{gather}
with the pion's decay constant $f_{\pi}$ and pion's mass $m_{\pi}$,
and the second term is the WZW term \cite{Son:2004tq,Son:2007ny} given by 
\begin{equation}
    \mathcal{L}_{\mathrm{WZW}} = -A^{\mathrm{B}}_{\mu} j_{\mathrm{GW}}^{\mu}.
    \label{effective_Lagrangian_GW}
\end{equation}
Here, 
the external \( \U(1)_{\mathrm{B}} \) gauge field \( A^{\mathrm{B}}_{\mu} \)  couples to \(\Sigma\) 
\cite{Son:2007ny} via the Goldstone-Wilczek current \cite{Goldstone:1981kk,Witten:1983tw}:
\begin{equation}
    j_{\mathrm{GW}}^{\mu} = -\frac{\epsilon^{\mu \nu \alpha \beta}}{24\pi^2} \mathrm{tr} \left( L_{\nu}L_{\alpha}L_{\beta} - 3\mathrm{i} e \partial_{\nu} \left[ A_{\alpha} Q(L_{\beta} + R_{\beta}) \right] \right) \,,
    \label{eq:GW}
\end{equation}
with \( A^{\mathrm{B}}_{\mu} = (\mu_{\mathrm{B}}, \bm{0}) \) 
and the left-invariant and right-invariant Maurer-Cartan 1-forms \( L_{\mu} \equiv \Sigma \partial_{\mu} \Sigma^\dagger \) and \( R_{\mu} \equiv \partial_{\mu} \Sigma^\dagger \Sigma \), 
respectively.

In particular, 
the contribution from the first term 
in eq.~(\ref{eq:GW}) to 
the action corresponding to the WZW term in eq.~(\ref{effective_Lagrangian_GW}) is
\begin{eqnarray}
    S_{\rm WZW}^{(1)} 
    &=&  
    \frac{1}{24\pi^2}
     \int_{M_4} d^4 x \,  
    \epsilon^{\mu \nu \alpha \beta}
    A_\mu^{\rm B}
    \mathrm{tr} \left( L_{\nu}L_{\alpha}L_{\beta}\right) \nonumber\\
    &=& -  \frac{\mu_{\rm B}}{24 \pi^2}  \int_{M_4} d^4 x \,
    \epsilon^{ijk} {\rm tr} [(\Sigma^\dagger \partial_i \Sigma)
    (\Sigma^\dagger \partial_j \Sigma)
    (\Sigma^\dagger \partial_k \Sigma)
    ] \nonumber\\
    &=&  \mu_{\rm B}   \int_{M_4} d^4 x \, {\cal B} \nonumber\\
      &=&  \frac{1}{24 \pi^2} 
    \int_{M_4} A_{\rm B} \wedge {\rm tr} [(\Sigma d \Sigma^\dagger)^3] 
    = -  \frac{1}{24 \pi^2} 
    \int_{M_4} A_{\rm B} \wedge {\rm tr} [(\Sigma^\dagger d \Sigma)^3] 
    \label{eq:WZW-1}
\end{eqnarray}
with $A^{\rm B} = A^{\rm B}_{\mu} dx^{\mu}$. 
Here, 
the baryon number density ${\cal B}$ is defined by
\begin{eqnarray}
    {\cal B} \equiv 
    \frac{1}{24\pi^2} \epsilon^{ijk}\, 
{\rm tr}\left( L_i L_j L_k \right)
= - 
    \frac{1}{24 \pi^2}
     \epsilon^{ijk} {\rm tr} [(\Sigma^\dagger \partial_i \Sigma)
    (\Sigma^\dagger \partial_j \Sigma)
    (\Sigma^\dagger \partial_k \Sigma)],
\end{eqnarray}
and its 
the integration over the three dimensional space 
gives the baryon number
\begin{eqnarray}
    N_{\rm B} \equiv \int d^3 x {\cal B} \in \pi_3[\SU(2)] \simeq {\mathbb Z}.
    \label{eq:def_baryon_number}
\end{eqnarray}

The terms in eq.~(\ref{eq:total-L})
are at the leading order 
${\cal O}(p^2)$ of a modification of the standard power counting scheme of ChPT 
 \cite{Brauner:2021sci}\footnote{
Here, we adopt a modification of the standard power counting scheme of ChPT as presented in ref.~\cite{Brauner:2021sci}:
    $\partial_{\mu} \,,
    m_{\pi} \,,
    A_{\mu} = \mathcal{O}(p^1)$ 
    and $A^{\mathrm{B}}_{\mu} = \mathcal{O}(p^{-1})$.
In this power-counting scheme,
eq.~(\ref{effective_Lagrangian_GW}) is of order $\mathcal{O}(p^2)$ and is consistent with eq.~(\ref{ChPT_with_B}).
It is significant to note that $\mu_{\mathrm{B}}$ only manifests in the WZW term of eq.~(\ref{effective_Lagrangian_GW}),
which allows us to attribute a negative power counting to $\mu_{\mathrm{B}}$.
}.
At the leading order,
the gauge field is nondynamical due to its kinetic term being of order $\mathcal{O}(p^4)$. 
For some of the solitons discussed in this paper, the $\mathcal{O}(p^4)$ terms may be needed for their stability, but the stability is not the main issue of this paper.

\section{Topological solitons in strong magnetic field in QCD} \label{sec:topo-solitons}

In this section, we introduce topological solitons in QCD matter in magnetic field: 
chiral solitons, domain-wall Skyrmions, neutral pion strings, pancake solitons(Skyrmions) and holes on a soliton.

\subsection{Chiral solitons}
We note that our effective theory admits a parallel stack of the sine-Gordon soliton expanding perpendicular to the external magnetic field,
which is called the CSL.
This state is stable under a sufficiently large magnetic field, as shown in ref.~\cite{Son:2007ny}.
If we consider the case of 
no charged pions 
$\Sigma_0 = \rme^{\rmi \tau_3\chi_3}$, the effective Lagrangian  reduces to 
\begin{eqnarray}
\mathcal{L}
= \frac{f_{\pi}^2}{2}(\del_{\mu}\chi_3)^2
    - f_{\pi}^2m_{\pi}^2(1-\textrm{cos}\chi_3)
  + \frac{e\mu_{\textrm{B}}B_z}{4\pi^2}\del_z\chi_3
\label{eq:lagrangian_without_charged_pions}
\end{eqnarray}
where, 
without loss of generality, we have 
oriented the uniform external magnetic field along the $z$-axis: $\bm{B}=(0,0, B_z)$
(with $F_{12} = - B_z$)
.

The ordinary QCD vacuum corresponds to $\chi_3=0$.
However, the third term in eq.~(\ref{eq:lagrangian_without_charged_pions}) modifies the ground state of QCD at finite $\mu_{\textrm{B}}$ and $B_z$.
The anticipated time-independent neutral pion background $\chi_3$ is obtained by minimization of the energy functional.
The static Hamiltonian depending only on the $z$ coordinate is given by
\begin{gather}
    \mathcal{H}
    = 
    \frac{f_{\pi}^2}{2}(\del_z\chi_3)^2
    + f_{\pi}^2m_{\pi}^2(1-\cos \chi_3)
    - \frac{e\mu_{\textrm{B}}B_z}{4\pi^2}\del_z\chi_3 \,. \label{eq:sine-Gordon_Hamiltonian_with_topo}
\end{gather}
We note that the last term in eq.~(\ref{eq:sine-Gordon_Hamiltonian_with_topo}) is the first derivative term $\del_z\chi_3$ 
which is a topological term.
Then, the configuration of the ground state will have a nontrivial $z$-dependence.
In order to determine the static configuration of $\chi_3$,
let us solve the EOM of eq.~(\ref{eq:lagrangian_without_charged_pions}).
The equation of motion for such a one-dimensional configuration $\chi_3(z)$ then reads
\begin{gather}
    \del_z^2\chi_3 = m_{\pi}^2\sin \chi_3 \label{eq:Jacobi-amplitude} \,.
\end{gather}
A single soliton solution is 
\begin{gather}
    \chi_3^{\rm single} = 4\tan^{-1}e^{m_{\pi}(z-Z)},  
\end{gather}
where $Z$ is the position of the center 
(or the translational modulus).
The tension, that is 
the energy density per unit area, can be analytically obtained:
\begin{align}
    E 
    = 8m_{\pi}f_{\pi}^2 - \frac{e\mu_{\textrm{B}}B_z}{2\pi} 
    \label{energy_of_DW} \,.
\end{align}

When $E<0$,
the sine-Gordon soliton is energetically more stable than the QCD vacuum ($\chi_3$=0).
The critical magnetic field at which the transition happens is 
\begin{gather}
    B_{\textrm{c}} = \frac{16\pi m_{\pi}f_{\pi}^2}{e\mu_{\textrm{B}}} \label{B_CSL} \,.
\end{gather}
The soliton carries the baryon number density 
\begin{eqnarray}
  b = \frac{eB_z}{2\pi}    
  \label{eq:baryon_number}
\end{eqnarray}
per a unit surface area.
The second term in eq.~(\ref{energy_of_DW}) is thus 
$-\mu_{\rm B}b$.

General solutions containing charged pions  can be obtained 
from $\Sigma_0$
by an $\SU(2)_{\textrm{V}}$ transformation,
\begin{gather}
    \Sigma = g\Sigma_0 g^{\dag} 
    = \exp(i \chi_3^{\rm single} g \tau_3 g^{\dag}) 
    \quad g \in \SU(2)_{\textrm{V}}   
    \label{general_sol} \,.
\end{gather}
Since $g$ in eq.~(\ref{general_sol}) 
is redundant with respect to 
a $\U(1)$ subgroup generated by $\tau_3$,
it takes a value in a coset space, 
\begin{eqnarray}
\SU(2)_{\textrm{V}}/\U(1)\simeq \mathbb{C}P^1 
\simeq S^2.
\end{eqnarray}
Together with the translational modulus $Z$, 
the single  sine-Gordon soliton has the moduli 
    $\mathcal{M} \simeq \mathbb{R} \times \mathbb{C}P^1$.
Such a soliton with non-Abelian moduli 
is called a non-Abelian sine-Gordon soliton 
\cite{Nitta:2014rxa,Eto:2015uqa,Nitta:2015mma,Nitta:2015mxa,Nitta:2022ahj}.
We now parameterize the $\mathbb{C}P^1$ moduli 
by the scalar triplet $\bm{n}=(n_1,n_2,n_3)$, 
satisfying 
\begin{gather}
    {\bm n}\cdot {\bm n} = 1 \,, \quad\quad
    g \tau_3 g^{\dag}
    = {\bm \tau}\cdot {\bm n} \equiv {\mathfrak n} \,.
\end{gather}
In terms of $\bm n$,
eq.~(\ref{general_sol}) can be rewritten as
\begin{gather}
    \Sigma = \exp({\rmi} {\mathfrak n}\chi_3^{\rm single}    ) = \cos \chi_3^{\rm single}
    +{\rmi} {\mathfrak n} \sin\chi_3^{\rm single}
    \label{general_sol_Sigma} \,.
\end{gather}

\subsection{Domain-wall Skyrmions}\label{sec:dwsk}
In this subsection, we introduce 
domain-wall Skyrmions.
In the previous papers \cite{Eto:2023lyo,Eto:2023wul}, 
topological lumps are constructed in the soliton effective theory 
(in a single-soliton approximation 
\cite{Eto:2023lyo}, 
and multiple solitons 
\cite{Eto:2023wul})
 as domain-wall Skyrmions.
 Here, we consider a single soliton 
 for our purpose to reveal spin statistics.
To this end, let us construct the effective theory 
on a single soliton. 
In terms of the $\rm O(3)$ vector $\bm n$ and single soliton profile $\chi_3^{\rm single}$, 
the Maurer-Cartan one forms
$L_k$ and $R_k$ can be expressed as
\begin{gather}
    L_k = - \rmi \mathfrak{n}\del_k\chi_3^{\rm single} -\frac{\rmi}{2} \del_k \mathfrak{n} \sin\left(2\chi_3^{\rm single}\right) +\frac{1}{2}[\mathfrak{n},\del_k\mathfrak{n}]\sin^2\chi_3^{\rm single} \,, \\
    R_k = - \rmi \mathfrak{n}\del_k\chi_3^{\rm single} -\frac{\rmi}{2} \del_k \mathfrak{n} \sin\left(2\chi_3^{\rm single}\right) - \frac{1}{2}[\mathfrak{n},\del_k\mathfrak{n}]\sin^2\chi_3^{\rm single} \,,
\end{gather}
respectively.
Since $\mathfrak n$ does not depend on $z$,
the second and third terms in $L_3$ and $R_3$ vanish.
Then, by using the moduli approximation 
(or Manton's approximation) \cite{Manton:1981mp,Eto:2006uw,Eto:2006pg},
we obtain the effective Lagrangian for 
a single soliton at ${\cal O}(p^2)$ as
\begin{eqnarray}
&&  \mathcal{L}_{\textrm {DW}} = 
        - 8m_{\pi}f_{\pi}^2 + \frac{e\mu_{\textrm{B}}B_z}{2\pi} +
     \mathcal{L}_{\textrm{norm}} 
     + \mathcal{L}_{\textrm{WZW}} ,
     \label{EFT_lag}\\
&&    \mathcal{L}_{\textrm{norm}}
    = \frac{4f_{\pi}^2}{3m_{\pi}} {\cal D}_\alpha {\bm n}\cdot {\cal D}^\alpha {\bm n}  
   \label{EFT_of_kinetic} \,,
    \quad \\
&&    \mathcal{L}_{\textrm{WZW}}
    = - 2\mu_{\textrm{B}}q
    +\frac{e \mu_{\textrm{B}}}{2\pi} \epsilon^{03jk}\del_j[A_k(1-n_3)]  ,
    \label{EFT_of_GW} \quad \quad
\end{eqnarray}
where the covariant derivative for the $\rm O(3)$ vector is defined by
\begin{equation}
    {\cal D}_\alpha {\bm n} = \del_\alpha {\bm n} -eA_\alpha ({\bm e_z}\times {\bm n})
\end{equation}
with the unit vector along $z$-axis ${\bm e}_z=(0,0,1)$.  
Here, $q$ is the lump (baby Skyrmion) 
topological charge density 
for $\pi_2({\mathbb C}P^1)$, defined by
\begin{eqnarray}
    q &\equiv&
     + \frac{1}{8\pi} \epsilon^{ij}{\bm n}\cdot (\partial_i {\bm n}\times\partial_j{\bm n}) \,.
    \label{eq:pi2} 
\end{eqnarray} 
Its integration gives
the topological lump number 
\begin{eqnarray}
     k &=& \int \rmd^2x\, q \in 
  \pi_2 ({\mathbb C}P^1) 
 \simeq \mathbb{Z}   .
 \label{eq:lump_num}
\end{eqnarray}
This topological term arises via the 
baryon number density ${\cal B}$ in the GW current 
\begin{eqnarray}
    q = - \frac{1}{2}\int_{-\infty}^{\infty} \rmd z~{\cal B}     .
    \label{lump_and_Skyrme}
\end{eqnarray}
Due to eqs.~(\ref{EFT_of_GW}) and (\ref{lump_and_Skyrme}),
the topological lump charge 
$k \in \pi_2(\mathbb{C}P^1)$ is related to the baryon number 
$N_{\rm B} \in \pi_3[\SU(2)]$ (topological charge 
of Skyrmions in the bulk) by
\begin{eqnarray}
    N_{\rm B} = \int \rmd^3x\, {\cal B} = -2\int \rmd^2x\, q 
    = \,- 2 k.\label{eq:NB}
\end{eqnarray}
Thus, $N_{\rm B}$ inside the soliton world-volume
is quantized in an even integer.

\begin{figure}[t]
\begin{center}
\begin{tabular}{c|c}
\includegraphics[width=8cm]{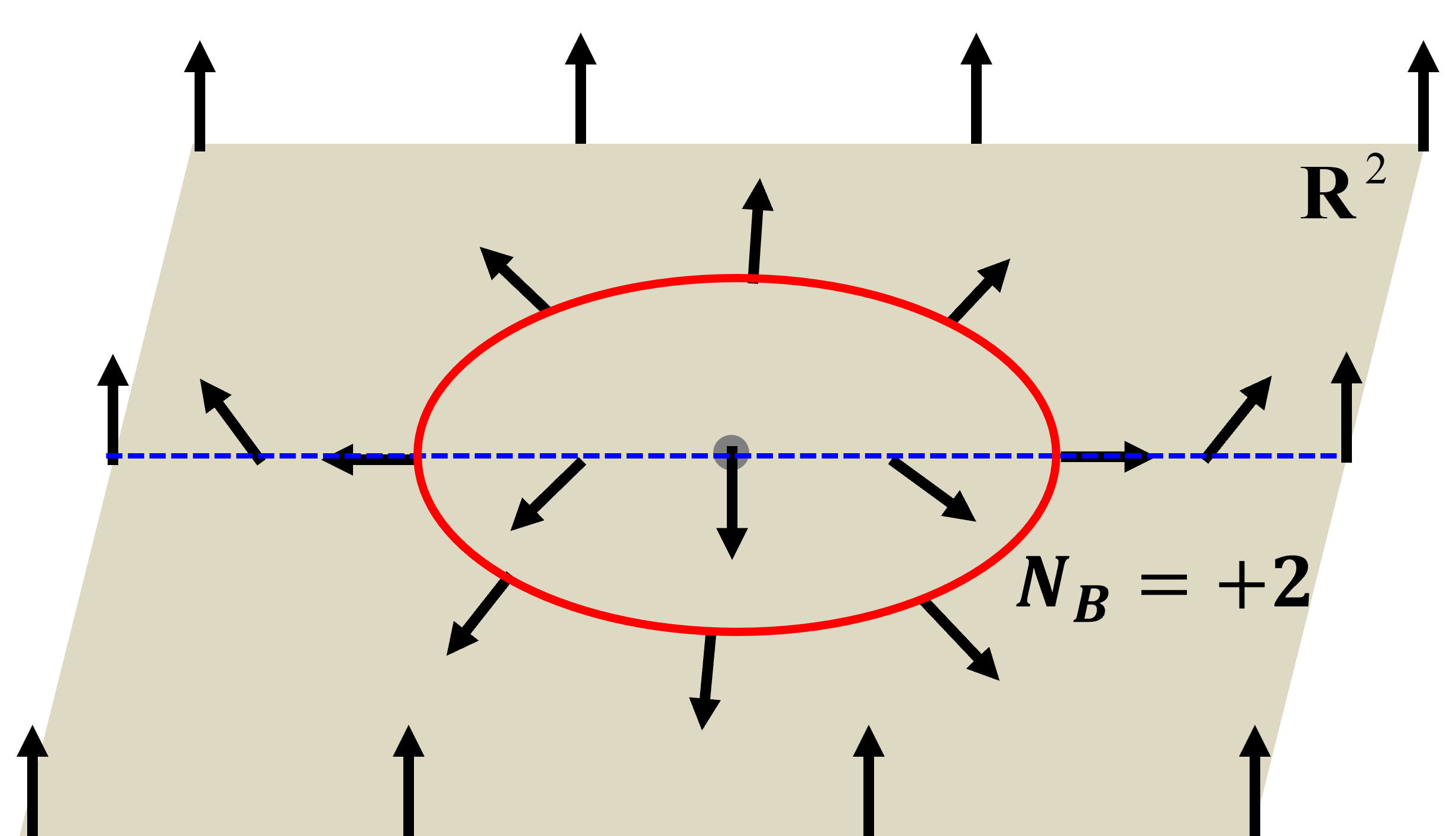} 
&
\includegraphics[width=8cm]{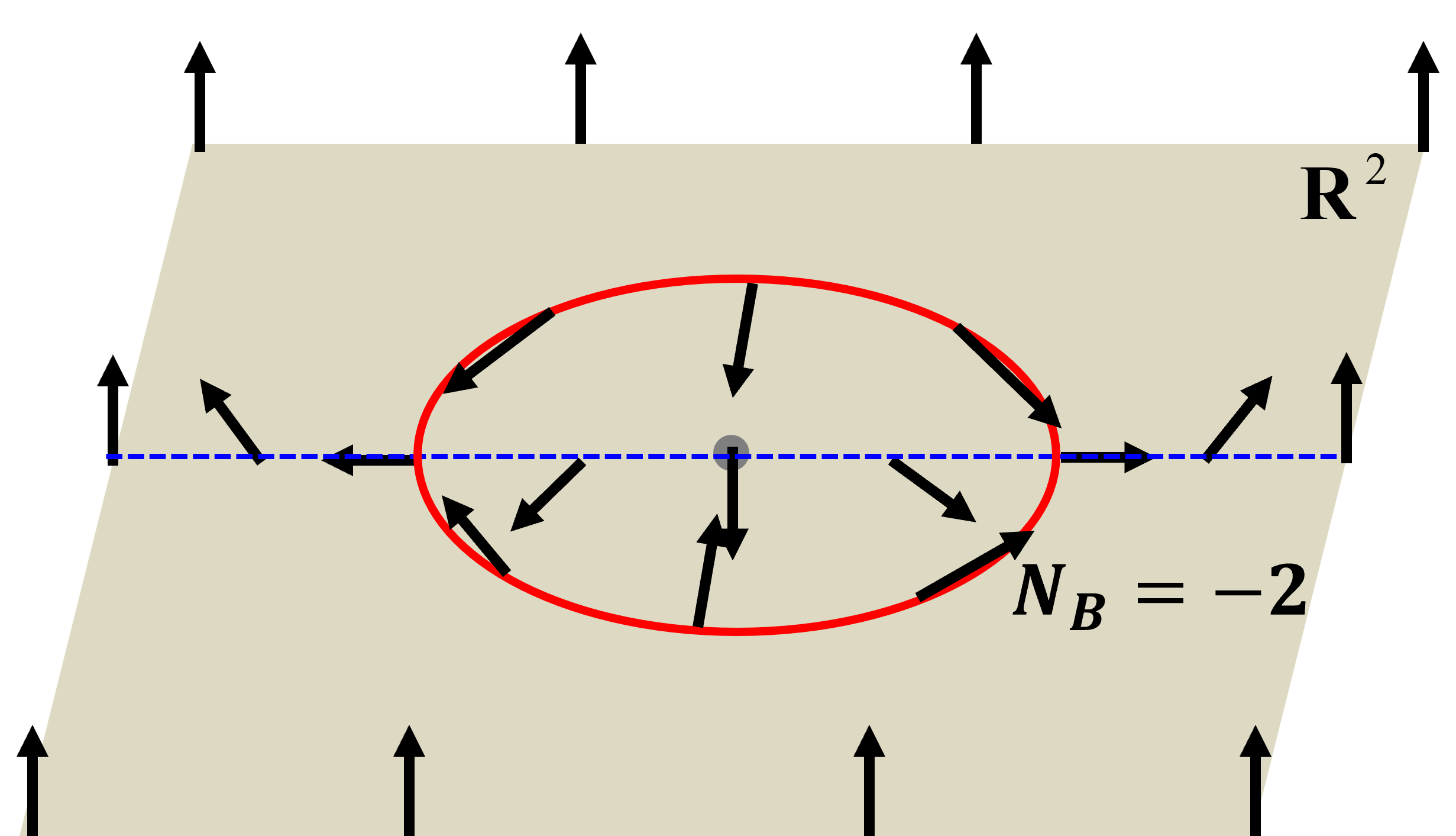} \\
(a) & (b)
\end{tabular}
\caption{Domain-wall Skyrmions. 
The topological lump number, 
vortex winding number, 
and baryon number 
are  
(a) $k=-1$,  $\nu =+1$ and $N_{\rm B} = + 2$, 
and 
(b) $k=+1$, $\nu=-1$ and $N_{\rm B} = + 2$.
The arrows denote the $S^2$ moduli on the soliton.
}
\label{fig:dwsk}
\end{center}
\end{figure}
Here, 
let us consider an axisymmetric  ansatz for 
lumps of the form
\begin{equation}
    {\bm n} = \left( \cos(\nu \varphi + \gamma)\sin f(\rho),
    \sin(\nu \varphi + \gamma)\sin f(\rho), \cos f(\rho)\right)
    \label{eq:skyrmion_ansatz}
\end{equation}
with the polar coordinates $(\rho, \varphi)$ 
in the $(x,y)$ plane, where $\nu\in \mathbb{Z}$ denotes a winding number, $\gamma\in[0,2\pi)$ is a constant describing the internal orientation (modulus) of the lump, and $f(\rho)\in[0,\pi]$ is a monotonic function satisfying the boundary conditions 
$\{f(0)=\pi, f(\infty) =0\}$.
By substituting the ansatz in eq.~(\ref{eq:skyrmion_ansatz}) into the topological charge 
in eq.~(\ref{eq:lump_num}), one obtains
\begin{equation}
    k =-\frac{\nu}{2}\left[\cos f(\rho)\right]_{\rho=0}^{\rho=\infty} = -\frac{\nu}{2}\left[n_3 (\rho)\right]_{\rho=0}^{\rho=\infty} .
    \label{eq:charge}
\end{equation}
In fig.~\ref{fig:dwsk}, we illustrate 
configurations of domain-wall Skyrmions.
These configurations 
 have the topological lump charges 
 and winding numbers 
(a) $k =-1$, $\nu = +1$
and (b)  $k =+1$, $\nu = -1$, respectively   
from eqs.~(\ref{eq:skyrmion_ansatz}) and 
(\ref{eq:charge}),
and thus the baryon numbers 
(a) $N_{\rm B} =+2$ 
and (b)  $N_{\rm B} =-2$ 
from eq.~(\ref{eq:NB}).
The configuration (b) is stable 
because of the quantization condition 
discussed below in eq.~(\ref{eq:quantization}). 
Although the configuration (a) is unstable for a nondynamical gauge field, it is stable once 
a gauge kinetic term is taken into account 
\cite{Amari:2024adu}.


\subsection{Neutral pion strings}\label{sec:pion-string}

In the next subsection, we explain that 
a soliton can be cut 
along a curve where the charged pions are excited.
To understand this, in this subsection, 
let us first consider 
a neutral pion vortex string.\footnote{
Pion strings at high temperature 
are discussed in a linear sigma model ~\cite{Zhang:1997is,Berera:2016vhw}.
}
The pion field can be expressed 
in terms of two complex scalar fields 
$(\phi_1, \phi_2)$ as
\begin{eqnarray}
    \Sigma = 
    \left(
\begin{array}{cc}
    \phi_1  & -\phi_2^* \\
    \phi_2 & \phi_1^* 
    \end{array}
    \right)
\end{eqnarray}
with a constraint $\det \Sigma = |\phi_1|^2 + |\phi_2|^2 = 1$ for 
$\SU(2)$. 
Here,
$\phi_1$ describes the neutral pion 
whereas $\phi_2$  does the charged pion.
In the chiral limit $m_{\pi} = 0$, 
the neutral pion can have a global vortex. 
Let us consider a straight neutral pion vortex string 
along the $x$-axis.
A straight axially symmetric neutral pion string is given by
\cite{Gudnason:2014hsa,Gudnason:2014jga,Gudnason:2016yix,Nitta:2015tua,Qiu:2024zpg}
\begin{eqnarray}
    \Sigma 
&=& 
\left(
\begin{array}{cc}
    f(r) e^{i\theta}  & -\sqrt{1-f(r)^2} e^{-i\phi(x)} \\
    \sqrt{1-f(r)^2} e^{i\phi(x)}& f(r) e^{-i\theta} 
    \end{array}
    \right)     \nonumber\\
&=& 
e^{\frac{i}{2} (\theta - \phi(x))\tau_3}
    \left(
\begin{array}{cc}
    f(r)   & -\sqrt{1-f(r)^2} \\
    \sqrt{1-f(r)^2} & f(r) 
    \end{array}
    \right)
e^{\frac{i}{2} (\theta + \phi(x)) \tau_3} 
    \label{eq:pion-string}
\end{eqnarray}
with the cylindrical coordinates $(r,\theta,x)$.
The profile function $f$ satisfies the boundary conditions 
$f(0) =0$ and $f(\infty) =1$. 
At the center of the vortex, the charged pions  
$\pi^+$ must appear 
with a $\rm U(1)$ phase as a modulus.
This $\rm U(1)$ modulus can be understood as 
a Nambu-Goldstone mode associated with 
the vector-like
$\rm U(1)$ symmetry generated by $\Sigma_3$ 
spontaneously broken 
in the vicinity of the vortex, 
as can be seen from the last expression 
in eq.~(\ref{eq:pion-string}).
This $\rm U(1)$ modulus is gauged by the electromagnetic 
symmetry, and thus it is a superconducting string \cite{Witten:1984eb}. 
It is important that 
if the phase $\phi(x)$ of the charged pion 
varies from $0$ to $2\pi$ along the string, 
it carries the baryon number $N_{\rm B} =1$
\cite{Gudnason:2014hsa,Gudnason:2014jga,Gudnason:2016yix,Nitta:2015tua}.

\begin{figure}[t]
\begin{center}
\includegraphics[width=9cm]{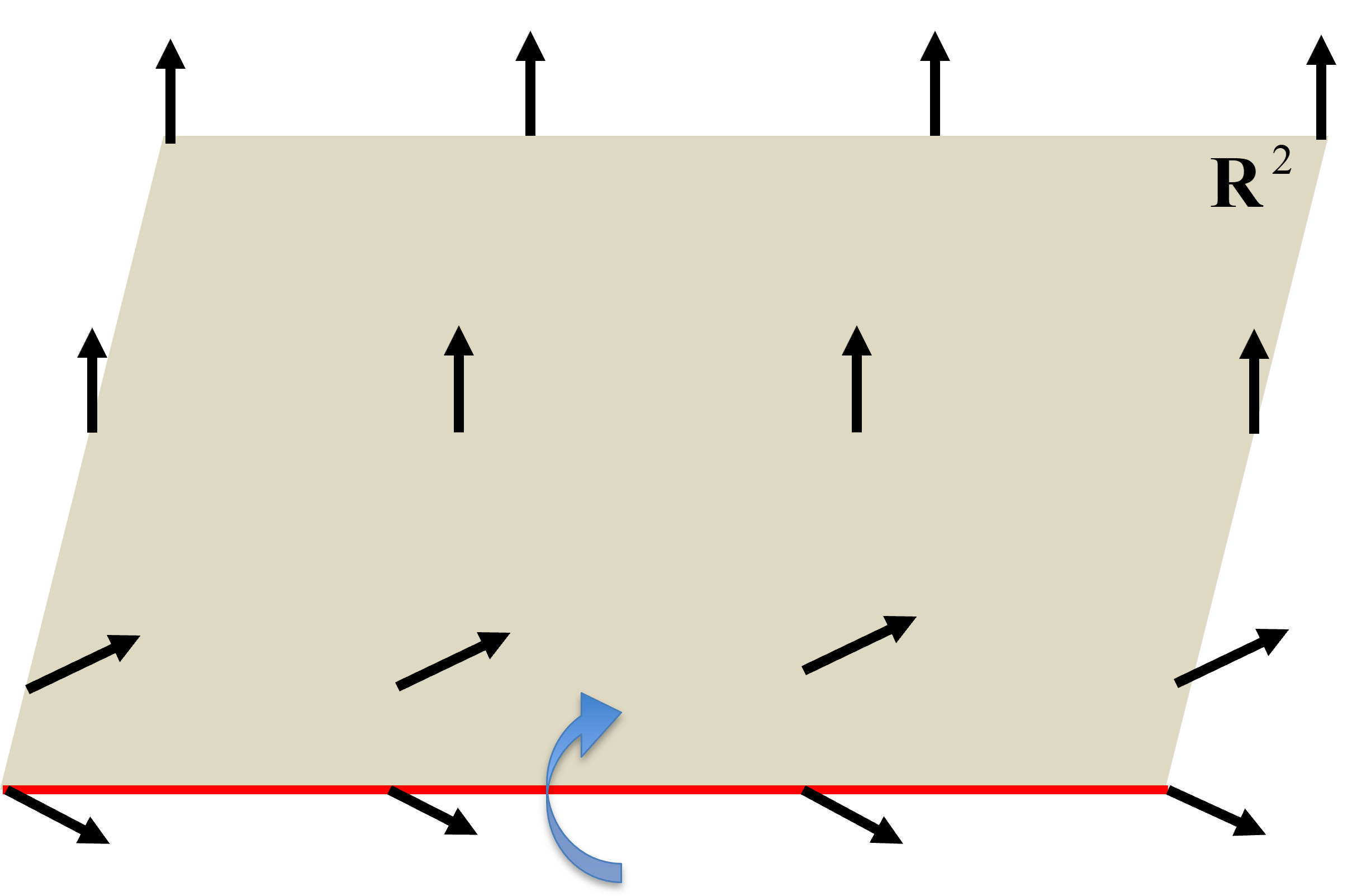}
\caption{The neutral pion string attached by a chiral soliton. 
The pion string is denoted by a red line.
The charged pions are present along the string, and thus it is a superconducting string. 
The ${\mathbb C}P^1$ moduli denoted by the arrows are present 
on the chiral soliton, and they are forced to lie on the plane 
at the boundary string.
}
\label{fig:pion-string}
\end{center}
\end{figure}
Let us turn on the pion mass $m_{\pi} \neq 0$. 
Then, the neutral pion string is attached by a neutral pion soliton, 
namely a chiral soliton 
\cite{Eto:2013hoa,Eto:2013bxa}.
This can be seen by modifying the ansatz in 
eq.~(\ref{eq:pion-string}) as
\begin{eqnarray}
    \Sigma = 
    \left(
\begin{array}{cc}
    f(r) e^{i\Theta(\theta)}  & -\sqrt{1-f(r)^2} e^{-i\phi(x)} \\
    \sqrt{1-f(r)^2} e^{i\phi(x)}& f(r) e^{-i\Theta(\theta)} 
    \end{array}
    \right) \label{eq:pion-string2}
\end{eqnarray}
with $\Theta$ a periodic function satisfying 
the boundary conditions 
$\Theta(0)=0$ and $\Theta(2\pi)= 2\pi$.  
At the large circle $r \to \infty$ encircling the vortex, 
the Lagrangian reduces to that of the sine-Gordon model:
\begin{eqnarray}
    {\cal L}_{r\to \infty} = \frac{f_\pi^2}{2}(\partial_\theta \Theta)^2 - \frac{f_\pi^2 m_\pi^2}{2}(1- \cos \Theta).
\end{eqnarray}
This implies that the neutral pion string 
is attached by a single semi-infinite chiral sine-Gordon soliton 
as axion strings, 
as illustrated in fig.~\ref{fig:pion-string}.
In other words, a chiral soliton can end on 
a neutral pion string.

The ${\mathbb C}P^1 \simeq S^2$ degrees of freedom live on 
the soliton indicated by the arrows in fig.~\ref{fig:pion-string}, as 
described in Sec.~\ref{sec:dwsk}.
On the other hand, only the $\rm U(1)$ modulus lives on the neutral 
pion string. They should be smoothly connected. 
This can be expressed by  representing the arrows  
lying parallel to the plane at the boundary.
We also can say that 
the soliton worldvolume can cut where 
the ${\mathbb C}P^1$ moduli lie on its worldvolume and the charged pions appear.

\subsection{Pancake solitons (Skyrmions)}

Here, we introduce a finite size soliton  
called a pancake soliton \cite{Son:2007ny}, 
which is nothing but an isolated Skyrmion 
in the presence of a magnetic field.
The finite size domain wall is a configuration
of the mesons and the magnetic field,
where the neutral pion has a winding number around 
the boundary of the domain wall,
and the phase of the charged pions $\pi^\pm $ has 
non-zero Aharonov-Bohm phase along the boundary.
The number of the finite size domain wall 
is quantized by the winding number of the neutral pion $\pi^0$ as well as the Aharonov-Bohm phase.

\begin{figure}[t]
\begin{center}
\begin{tabular}{cc}
\includegraphics[width=7cm]{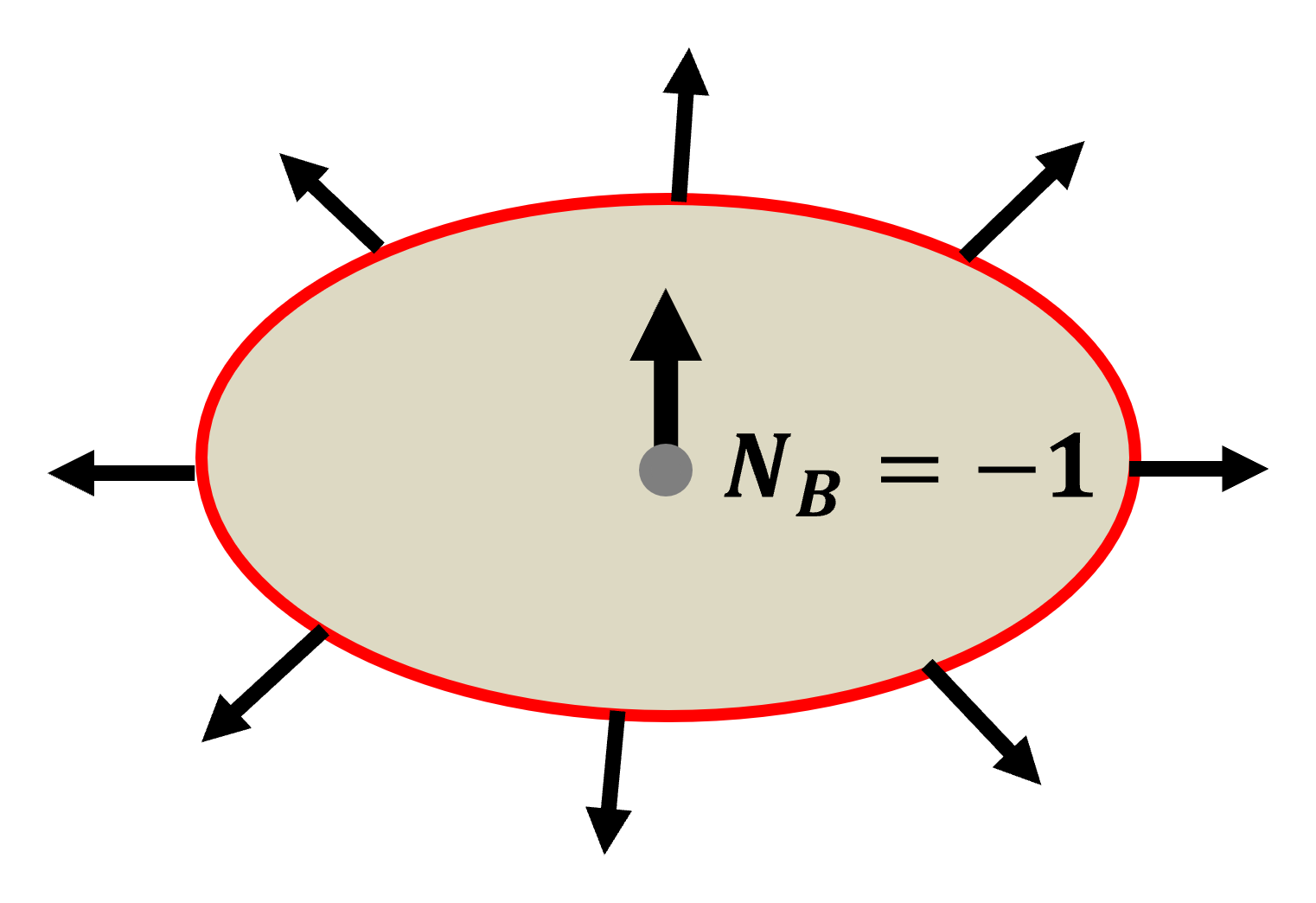} 
&
\includegraphics[width=7cm]{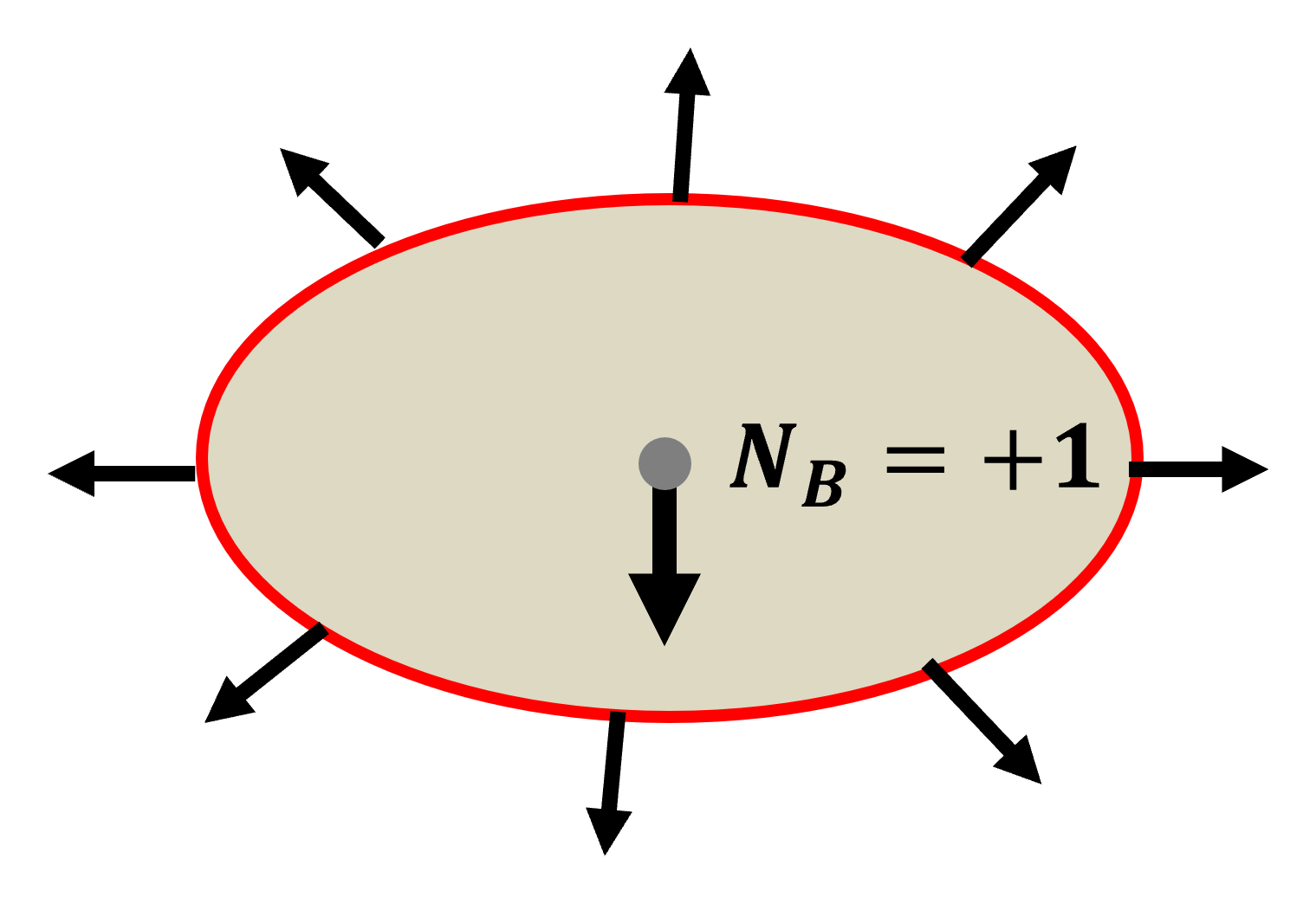}
\\ 
(a) & (b)\\ 
\\
\includegraphics[width=7cm]{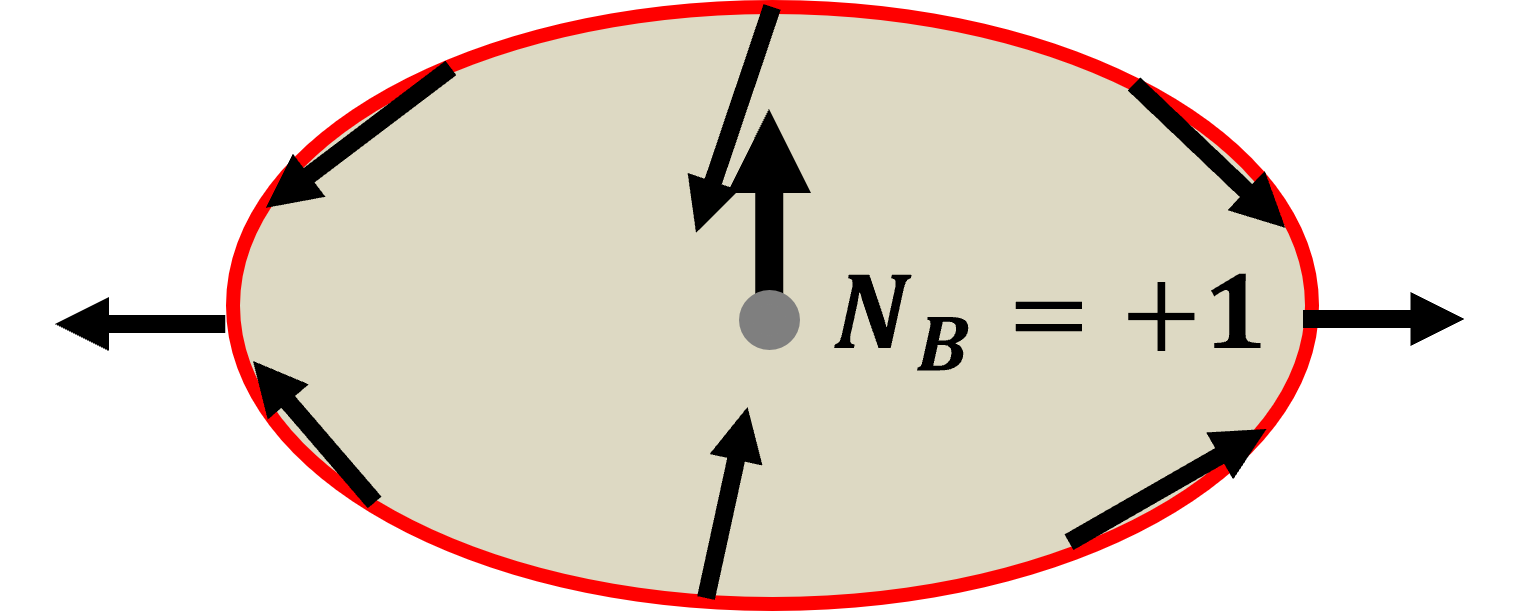} 
&
\includegraphics[width=7cm]{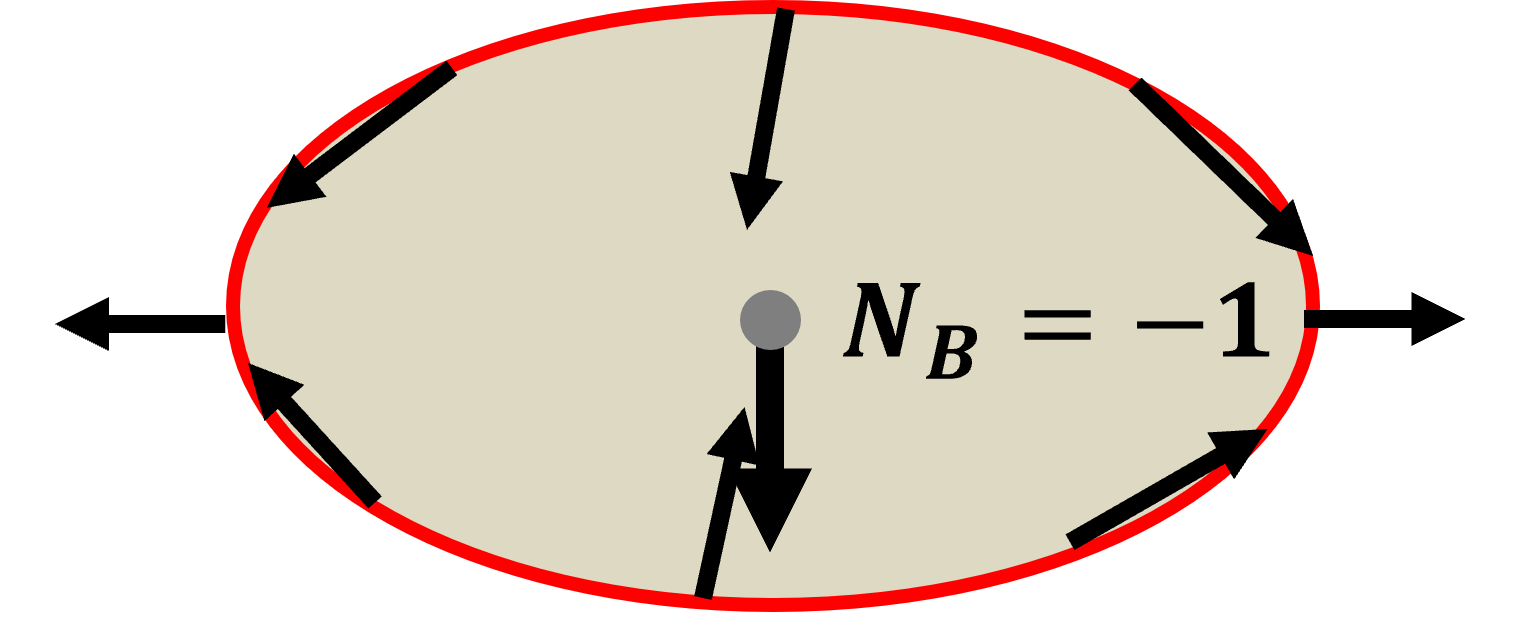}\\
(c) & (d)
\end{tabular}
\caption{Pancake solitons. 
The arrow denote the ${\mathbb C}P^1$ moduli ${\bm n}$ on the 
soliton effective theory.
The phase $\alpha$ of the neutral pion $\pi^0$ 
changes as $0 \to 2\pi$ from the bottom to top along the $z$-axis 
 (the lines penetrating the pancakes)  
in (a) and (c), while it changes as 
$0 \to -2\pi$ in (b) and (d). 
It passes 
through the north [south pole] and a point at the equator of the sphere $S^2$ 
for (a) and (c) [(b) and (d)], 
as indicated by the arrow $\uparrow$ ($\downarrow$) 
 at the centers of the pancakes.
 The charged pion 
 $\pi^\mp = \pi^1 \pm i \pi^2 \sim e^{\pm i\phi}$
 appear on the edges 
 of the pancakes.
The horizontal arrows also represent   the $\rm U(1)$ phase $\phi$ of $\pi^-$, 
winding along the edges with 
 the winding numbers (a) and (b) $\nu = +1$, (c) and (d) $\nu = -1$.
  The topological lump charges are 
(a) $k= +1/2$,  
(b) $k= -1/2$,  
(c) $k= -1/2$,  
(d) $k= +1/2$
from eq.~(\ref{eq:charge-pancake}). 
 The baryon numbers 
 (Skyrme numbers $\pi_3[\SU(2)] \simeq {\mathbb Z}$) are  
(a) $N_{\rm B} = - 1$,  
(b) $N_{\rm B} = + 1$,  
(c) $N_{\rm B} = + 1$,  
(d) $N_{\rm B} = - 1$. 
The configurations  
 (a) and (d) [(b) and (c)] can be rotated to each other, 
 for which 
(a) and (d) [(b) and (c)]
are energetically degenerated in the absence of an external magnetic field.
 When the magnetic field is applied, degeneracies are lifted and the sizes are quantized. 
 If the magnetic field is applied 
 along the positive (negative) $z$ direction, 
 (c) and (d) have
 lower (higher) energy than (b) and (a), respectively.
 If the gauge field is dynamical, then a magnetic field is induced for 
 (c) and (d)
 [(a) and (b)] 
 along the positive (negative) $z$ direction, as they behave as magnetic moments.
}
\label{fig:pancake}
\end{center}
\end{figure}

First, we review the baryon number of the finite size domain wall~\cite{Son:2007ny}. 
See fig.~\ref{fig:pancake}.
For concreteness, we assume that 
the finite size domain wall is a disk
parameterized by $ 0 \leq \rho \leq R$,
$z =0$ 
in the cylindrical coordinates 
$(\rho, \varphi, z)$.
It will be convenient to take 
the parameterization of the meson field as
$\Sigma = \sigma +  i \pi^a \tau^a $
with 
\begin{eqnarray}
    \sigma & = \cos \chi \cos \alpha , 
\label{240423.1426}
    \\ 
    \pi^0 & = \cos \chi \sin \alpha,
    \\ 
        \pi^1 & = \sin \chi \cos \phi, 
    \\ 
    \pi^2 & = \sin \chi \sin \phi.
\label{240423.1427}     
\end{eqnarray}
or in the matrix form, 
\begin{equation}
\Sigma
 = 
\mtx{
\sigma + i\pi^0 & i(\pi^1 - i \pi^2)  
\\ 
i(\pi^1 + i \pi^2)  & \sigma - i\pi^0   
}
 = 
\mtx{
e^{i\alpha}  \cos\chi 
 & i e^{- i  \phi }   \sin \chi \\ 
i e^{i \phi } \sin \chi   & e^{ - i\alpha }  \cos\chi }
.
\label{eq:parametrization_Sigma_pancake}
\end{equation}
The phase $\alpha$ of the neutral pion jumps by $\pm 2 \pi$ 
when it goes through inside the domain wall:
$\alpha (\rho  < R , \varphi, z = +\infty)
-  \alpha (\rho < R , \varphi, z = - \infty)
 = \pm 2\pi$
 and
$ \alpha (\rho >R , \varphi, z = +\infty)
- \alpha (\rho >R , \varphi, z = - \infty)
 = 0$.
 Here, the upper (lower) sign corresponds to the 
 upward (downward) magnetic field.
In this configuration, 
the $\pi^0$ meson has a winding number
around the boundary
$\int_{\cal C} d\alpha = \pm 2\pi$, where
${\cal C}$ is a loop around the boundary.
Since $\pi^0$ has a winding number around 
the boundary, we require 
$\chi (\rho = R, z=0) = \frac{\pi}{2} $
to avoid the singularity in $\Sigma$.
Therefore, the charged meson fields 
satisfy
$(\pi^1)^2 + (\pi^2)^2 = 1$ 
on the boundary.
This shows that the boundary of the pancake soliton 
is in the Higgs phase. 
It is thus a superconducting ring.

To minimize the energy, the charged pion fields 
$
 \pi^\pm = \pi^1 \mp i \pi^2   
 = i e^{\mp i\phi} \sin \chi
$
should satisfy 
\begin{equation}
  (\der_\mu \pm i e A_\mu) \pi^\pm  =0   
\quad \Leftrightarrow \quad
  A_\mu = \pm  i e^{-1} (\pi^\pm)^{-1} \partial_\mu \pi^\pm 
\end{equation}
on the boundary 
where $\chi =\pi/2$.
Therefore, the minus winding number of the charged pion $\pi^-$ 
along the boundary is equal to the magnetic flux 
through the pancake soliton,
\begin{equation}
    - \int_{S^1} 
    \der_\mu \phi dx^\mu 
    =
      \int_{S^1} \fr{i }{ \pi^{-}} 
     \der_\mu \pi^{-} dx^\mu 
     = 
     {-} \int_{S^1} A_\mu dx^\mu 
     = 
      e \int_{D^2} \bs{B} \cdot d \bs{S}
    =  e B_z S \in 2 \pi \bb{Z},\label{eq:quantization}
\end{equation}
where 
$S = \pi R^2$ 
is the area of the finite size domain wall.
Since this winding number is also topological but non-zero, we require 
$\chi (\rho =0 ) = 0 $ to avoid the singularity in $\Sigma$.
For $B_z >0$ ($B_z<0$), 
negative (positive) winding of the phase 
$\phi$ can meet the quantization condition 
in eq.~(\ref{eq:quantization}). 
The integer in eq.~(\ref{eq:quantization}) 
is nothing but the  baryon number $N_{\rm B}$.
In such a case, 
eq.~(\ref{eq:quantization}) can be rewritten by
\begin{eqnarray}
    S = \frac{2\pi }{e B_z} N_{\rm B} ,\quad
\mbox{or} \quad  R = \sqrt {\frac{2 N_{\rm B}}{eB_z}}    
\end{eqnarray}
where $N_{\rm B}>0$ ($N_{\rm B}<0$) for  $B_z >0$ ($B_z<0$).
It is important to note that 
the baryon number density 
$b \equiv N_{\rm B} / S = e B_z /2\pi$ 
coincides with that of 
an infinitely large chiral soliton in eq.~(\ref{eq:baryon_number}).

\begin{figure}[t]
\begin{center}
\begin{tabular}{cc}
\vspace{-0.2cm}
\includegraphics[width=6cm]{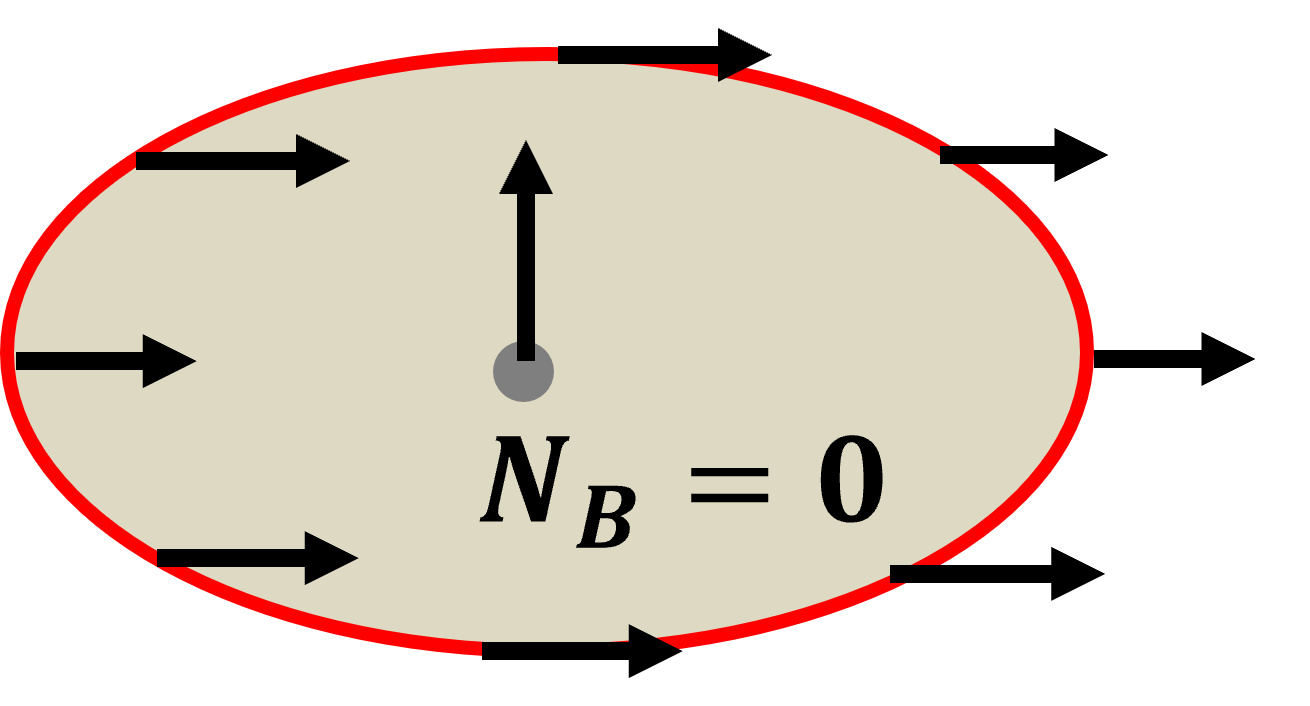} & \includegraphics[width=6cm]{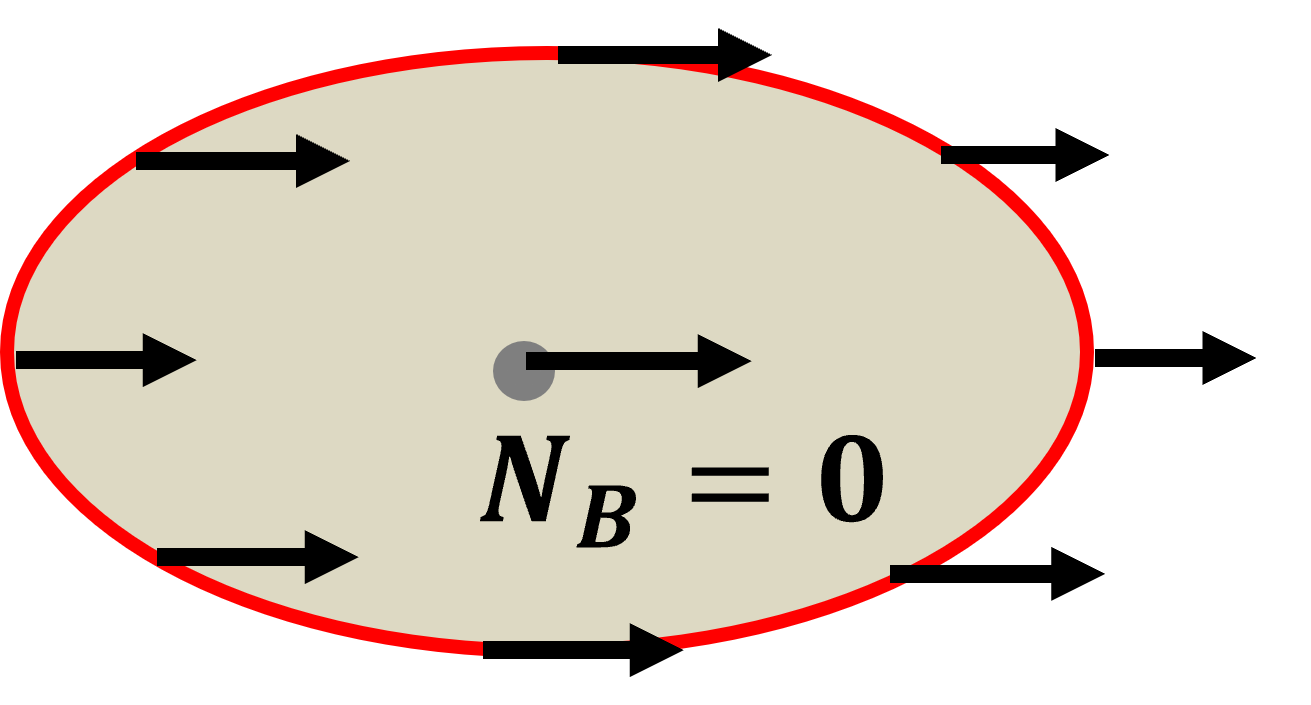} \\
(a) & (b)
\end{tabular}
\caption{
Nontopological pancake solitons.
(a) 
The phase of the neutral pion $\pi_0$ 
changes as $0 \to 2\pi$ from the bottom to top along the $z$-axis 
as fig.~\ref{fig:pancake}(a). 
However, the charged pion does not wind on the edge 
unlike fig.~\ref{fig:pancake}(a) and (b), 
resulting in $N_{\rm B}=0$.
(b) The configuration (a) energetically 
deforms to (b). Then, it is unstable.
Note that such a deformation is impossible 
for the topological pancake solitons in fig.~\ref{fig:pancake}.
\label{fig:pancake2}
}
\end{center}
\end{figure}

The pancake solitons can be constructed 
in the effective theory as follows.
Although the effective theory was originally constructed for an infinitely large soliton 
in sec.~\ref{sec:dwsk}, the soliton can end on a neutral pion string where the charged pions appear, as discussed in sec.~\ref{sec:pion-string}. Therefore, we can consider the effective theory of a chiral soliton of a finite size with a proper boundary condition.
A pancake soliton of the radius $R$ is 
the same with eq.~(\ref{eq:skyrmion_ansatz}) 
with the polar coordinates $(\rho, \varphi)$ 
($0 \leq \rho \leq R$)  
in the $(x,y)$ plane. 
$\nu\in \mathbb{Z}$ denotes a winding number 
at the boundary $\rho=R$, $\gamma\in[0,2\pi)$ is a constant describing the internal orientation, and $f(\rho)\in[0,\pi]$ is a monotonic function satisfying the boundary conditions 
$\{f(0)=0, f(R) =\pi\}$ or $\{f(0)=\pi, f(R) =0\}$.
By substituting the ansatz in eq.~(\ref{eq:skyrmion_ansatz}) into the topological charge 
in eq.~(\ref{lump_and_Skyrme}), one obtains
\begin{equation}
    k =-\frac{\nu}{2}\left[\cos f(\rho)\right]_{\rho=0}^{\rho=R} = -\frac{\nu}{2}\left[n_3 (\rho)\right]_{\rho=0}^{\rho=R} .
    \label{eq:charge-pancake}
\end{equation}
 
In fig.~\ref{fig:pancake}, we illustrate four possibilities 
of topologically nontrivial pancake solitons of a radius $R$ 
with the minimum topological charges.  
 The charged pions 
 $\pi^\pm = \pi^1 \mp i \pi^2$ appear 
 with proportional to $n_1 \mp i n_2$ 
 where $n_3=0$
 on the edges 
 of the pancakes, denoted by the red circles.  
 The $\rm U(1)$ phases have nontrivial windings $\nu$ 
 as denoted by the horizontal arrows; 
(a) $\nu = +1$, (b)  $\nu = +1$  
(c) $\nu = - 1$, (d) $\nu = - 1$.  
From eq.~(\ref{eq:charge-pancake})
the topological lump charges 
on the pancake soliton can be calculated as 
(a) $k= +1/2$,  
(b) $k= -1/2$,  
(c) $k= -1/2$,  
(d) $k= +1/2$, respectively,
and thus
they are half-quantized lumps, 
or sometimes called merons. 
From eq.~(\ref{eq:NB}) 
 the baryon numbers $N_{\rm B}$
 (the Skyrmion numbers $\pi_3[\SU(2)] \simeq {\mathbb Z}$) are  
(a) $N_{\rm B} = - 1$,  
(b) $N_{\rm B} = + 1$,  
(c) $N_{\rm B} = + 1$,  
(d) $N_{\rm B} = - 1$.

The configurations  
 (a) and (d) [(b) and (c)] can be rotated to each other by $\pi$ around the $x$ or $y$ axis. 
 If there is no magnetic field,  
the all configurations 
are energetically degenerated 
and their sizes are not quantized.
 In the presence of 
 the upward (downward) magnetic field,
 the degeneracy is lifted as follows:
 First, 
  in the presence the upward (downward) magnetic field,
 the  change of the phase of the neutral pion for (d) and (b) [(a) and (c)] are not energetically 
 preferred from the WZW term, 
 and thus they rotate to (a) and (c) 
 [(d) and (b)], respectively. 
Second, the pancake solitons 
can be regarded as superconducting rings, 
since 
the charged pions are localized around their edges.
Then, the configurations (c) and (d) are preferred compared 
with the configurations (b) and (a), respectively.\footnote{ 
If the gauge field is dynamical, 
these pancake solitons can be regarded as magnetic moments since an upward (downward) magnetic field is induced for (c) and (d) [(a) and (b)].
Thus, the magnetic moment prefers to be aligned to the external magnetic field.
}
The sizes of the configurations (c) and (d) are quantized by eq.~(\ref{eq:quantization}), 
or they can minimize the covariant derivative of the charged pions 
along the loop, 
while those of (a) and (b) are not.

We also have pancake solitons with 
the higher topological charges $N_{\rm B}$ 
whose surface area are proportional to $N_{\rm B}$ 
from 
the quantization condition 
in eq.~(\ref{eq:quantization}).
Such a behavior is one of droplets.

In fig.~\ref{fig:pancake2}, an example 
of topologically trivial pancake soliton 
with the winding number of charged pions 
at the edge is zero. 
Energetically the configuration (a) 
deforms to the one (b), 
and it can disappear after shrinking.
Note that such a deformation is impossible 
for the topological pancake solitons in fig.~\ref{fig:pancake}.

\subsection{Holes on a soliton}

\begin{figure}[t]
\begin{center}
\begin{tabular}{cc}
\includegraphics[width=7cm]{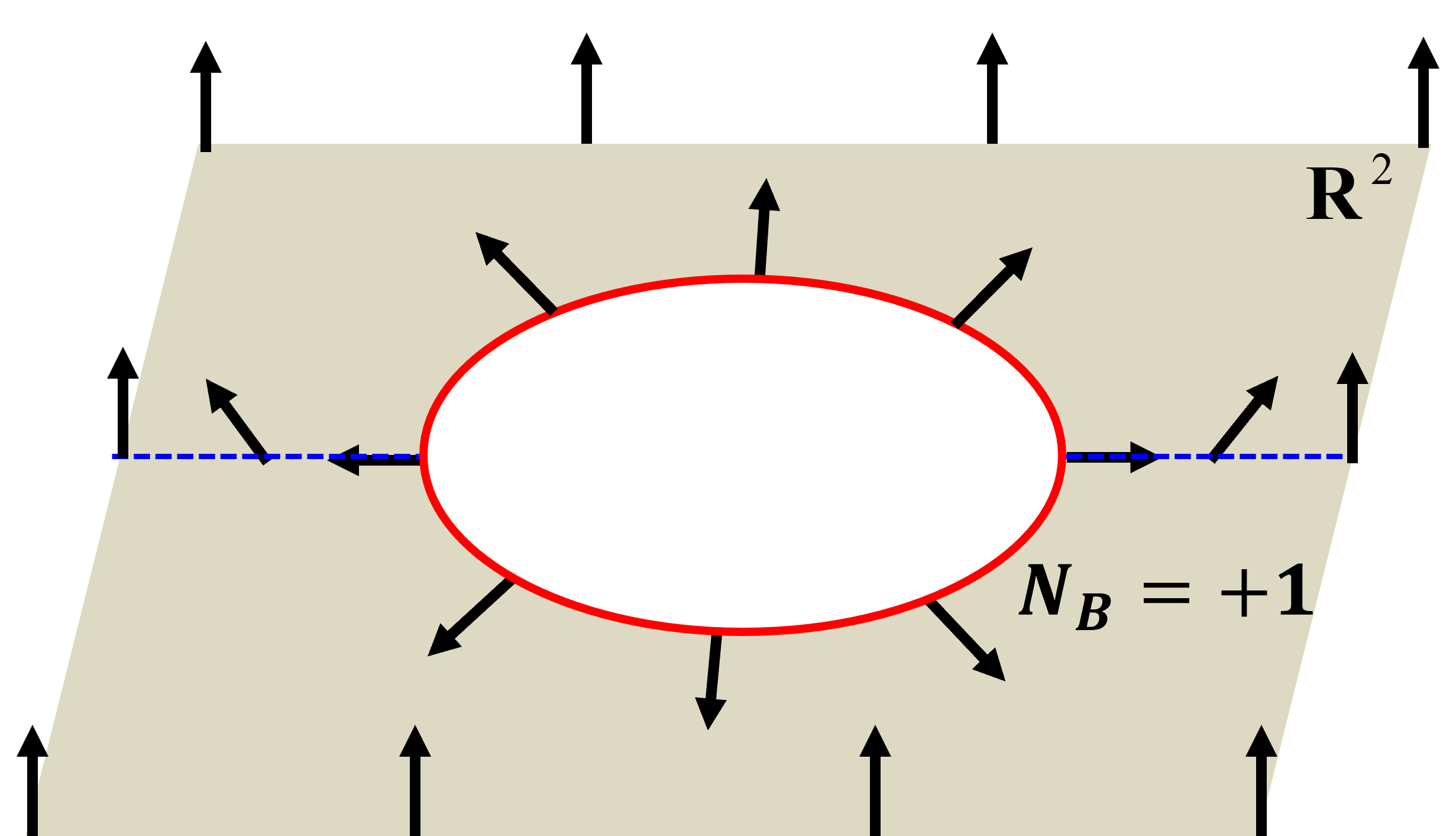} 
&
\includegraphics[width=7cm]{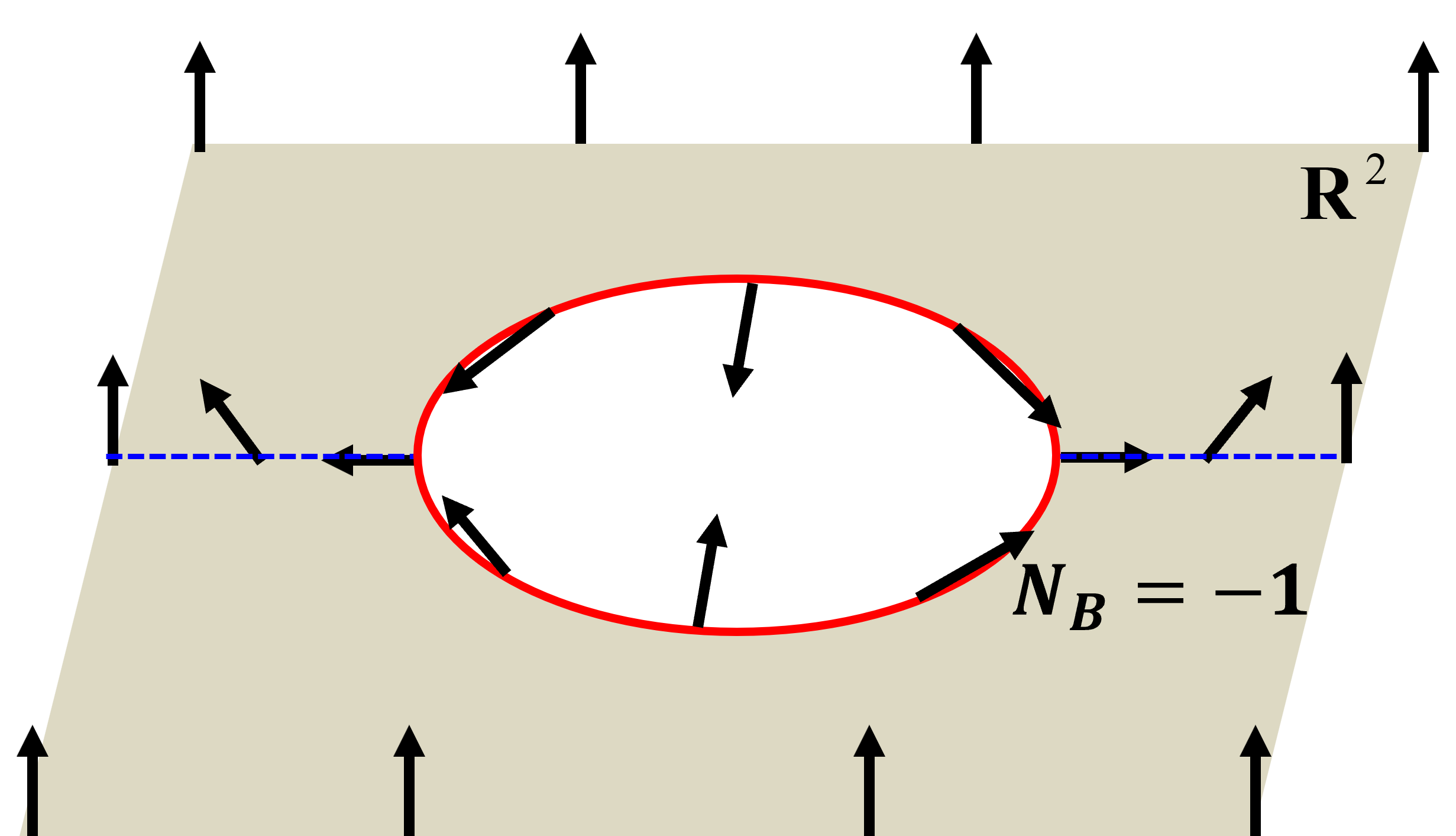}\\ 
(a) & (b)
\end{tabular}
\includegraphics[width=7cm]{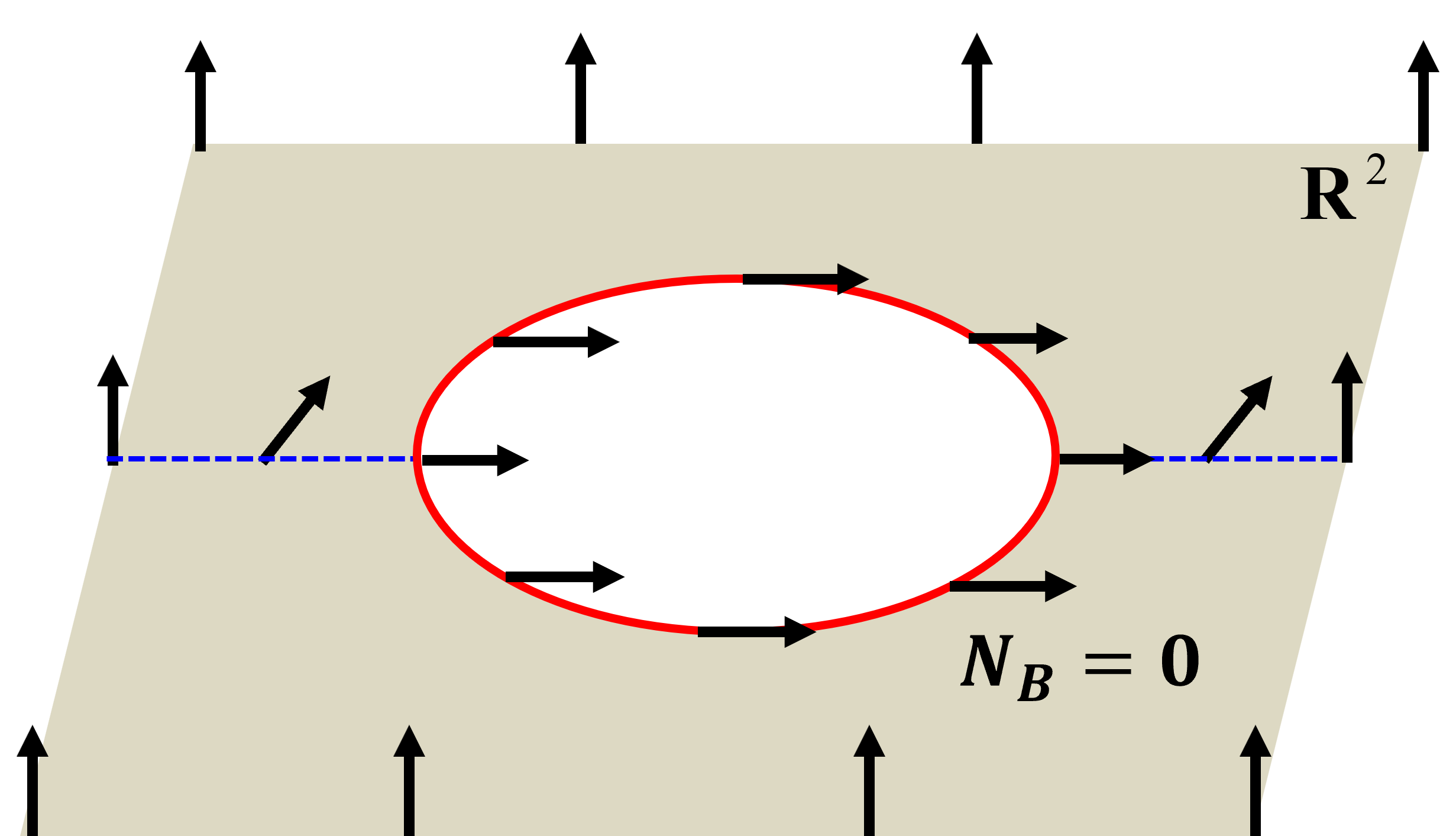} \\
(c)  
\caption{Holes on a soliton. 
The holes are bounded by 
closed $\pi^0$ strings 
on which 
the charged pions appear. 
The phase of the charged pions is represented by 
the horizontal arrows on the soliton.
Topologically nontrivial holes with 
the winding number (a) $\nu = +1$ and (b) $\nu = -1$. 
(c) A topologically trivial hole 
(with the winding number $\nu = 0$). 
The topological lump numbers are  
(a) $k = -1/2$, (b) $k = +1/2$, 
(c) $k = 0$.  
The baryon numbers are 
(a) $N_{\rm B} = +1$, (b) $N_{\rm B} = -1$, 
(c) $N_{\rm B} = 0$. 
In the presence of the upward(downward) magnetic field, 
the size of (b) [(a)] is  quantized. 
There also exists a hole with a higher winding number $\nu$ 
and $N_{\rm B} = - \nu$. 
The solitons with $\downarrow$ are energetically disfavored 
and are not drawn.
}
\label{fig:hole}
\end{center}
\end{figure}
We also can construct a hole on a single soliton, 
which was not discussed before.
The hole is bounded by 
a closed neutral pion string  
on which 
the charged pions appear. 
Again, the phase of the charged pions can have 
nontrivial winding along the closed loop.
A hole on the soliton of the radius $R$ is 
the same with eq.~(\ref{eq:skyrmion_ansatz}) 
with the polar coordinates $(\rho, \varphi)$ 
($R \leq \rho < \infty$)  
in the $(x,y)$ plane. 
Again, 
$\nu\in \mathbb{Z}$ denotes a winding number 
at the boundary $\rho=R$, $\gamma\in[0,2\pi)$ is a constant describing the internal orientation, and $f(\rho)\in[0,\pi]$ is a monotonic function satisfying the boundary conditions 
$\{f(R)=0, f(\infty) =\pi\}$ 
or $\{f(R)=\pi, f(\infty) =0\}$.
By substituting the ansatz in eq.~(\ref{eq:skyrmion_ansatz}) into the topological charge 
in eq.~(\ref{lump_and_Skyrme}), one obtains
\begin{equation}
    k =-\frac{\nu}{2}\left[\cos f(\rho)\right]_{\rho=R}^{\rho=\infty} = -\frac{\nu}{2}\left[n_3 (\rho)\right]_{\rho=R}^{\rho=\infty} .
    \label{eq:charge-hole}
\end{equation}
In fig.~\ref{fig:hole}, we illustrated 
holes 
with the winding numbers 
(a) $\nu =+1$, 
(b) $\nu =-1$, 
(c) $\nu =0$. 
From eq.~(\ref{eq:charge-hole})
the topological lump charges are 
(a) $k = -1/2$, (b) $k = +1/2$, 
(c) $k = 0$.  
Then, 
from eq.~(\ref{eq:NB}) 
 the baryon numbers $N_{\rm B}$ are 
(a) $N_{\rm B} = +1$, (b) $N_{\rm B} = -1$, 
(c) $N_{\rm B} = 0$.

In the presence of 
the upward(downward) magnetic field,  
the radius of the hole (b) [(a)] is quantized 
by eq.~(\ref{eq:quantization}) 
in the same way as pancake solitons, 
and thus it is stable as it is. 
The hole (a) [(b)]
 cannot  satisfy 
the quantization condition in 
eq.~(\ref{eq:quantization}) 
in 
the upward(downward) magnetic field, 
 is unstable to shrink, 
 and eventually becomes 
 singular unless 
 either 
 the Skyrme term or 
 dynamical gauge field is taken into account.
The nontopological hole (c) is unstable to shrink in any case, 
and will disappear.

\section{Surgeries of topological solitons}\label{sec:surgery}

In this section, we propose surgeries 
of a domain-wall Skyrmion and a chiral soliton lattice.

\subsection{Surgery of a domain-wall Skyrmion}
\begin{figure}[t]
\begin{center}
\includegraphics[width=10cm]{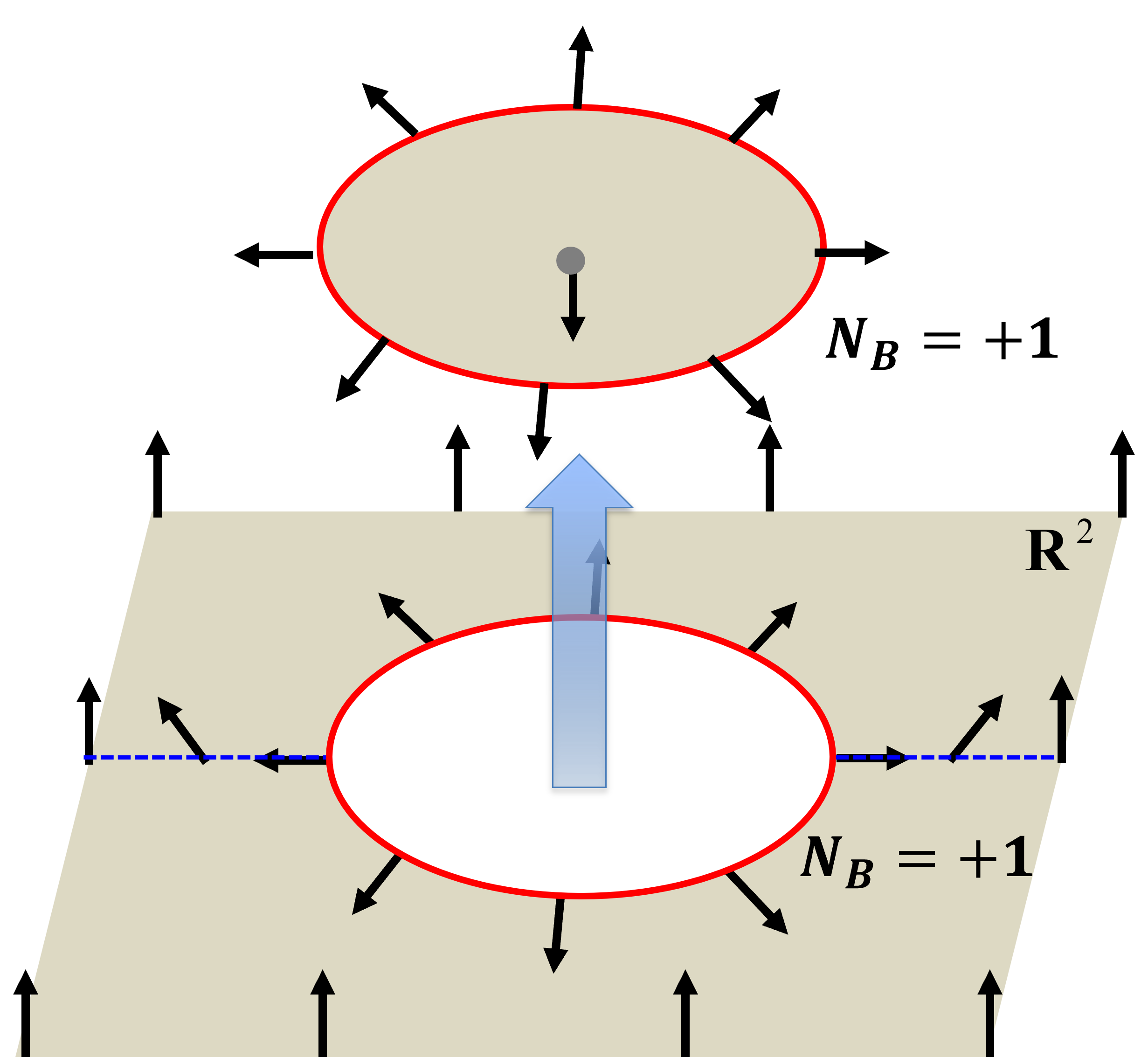}
\caption{Surgery of a domain-wall Skyrmion.
Under a surgery, one can cut 
a domain-wall Skyrmion 
along the red curve where charged pions are induced. 
Then, the domain-wall Skyrmion 
is decomposed into 
a pancake soliton 
of the baryon number 
$N_{\rm B} = 1$ 
bounded by a closed $\pi_0$ 
string, and a soliton with a hole 
of the baryon number 
$N_{\rm B} = 1$ 
bounded by a closed $\pi^0$ anti-string. 
The pancake soliton is one of fig.~\ref{fig:pancake}(b).
This does not satisfy 
the quantization condition
and rotates by $\pi$
to fig.~\ref{fig:pancake}(c). 
}
\label{fig:surgery1}
\end{center}
\end{figure}

From sec.~\ref{sec:pion-string},
one can understand that 
a soliton can be cut 
along a curve where the charged pions are excited.
Therefore, 
a domain-wall Skyrmion in fig.~\ref{fig:dwsk}(a)
can be cut along the red circle along which the charged pions 
are present, resulting in a surgery of decomposition  
into a soliton with a hole and a pancake soliton, 
as illustrated in fig.~\ref{fig:surgery1}.
The pancake soliton obtained in this way is one of fig.~\ref{fig:pancake}(b). 
Since  this configuration is not energetically preferred one, 
it rotates 
to one of fig.~\ref{fig:pancake}(c). 
Likewise, 
a domain-wall anti-Skyrmion in fig.~\ref{fig:dwsk}(b) can also be decomposed.

\begin{figure}
\begin{center}
\includegraphics[width=6cm]{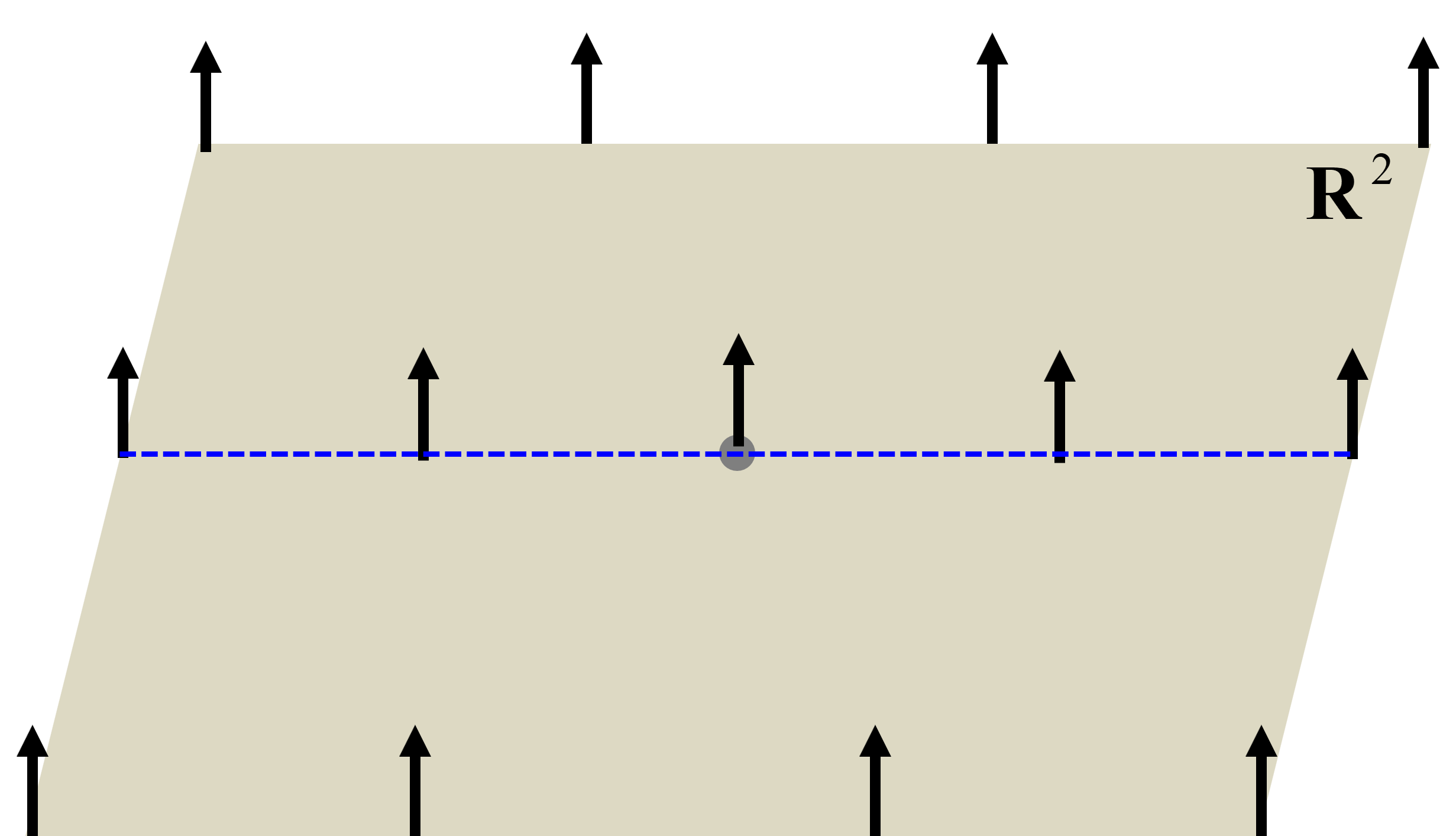}\\
(a)\\
\vspace{-1cm}
\begin{tabular}{cc}
\includegraphics[width=6.5cm]{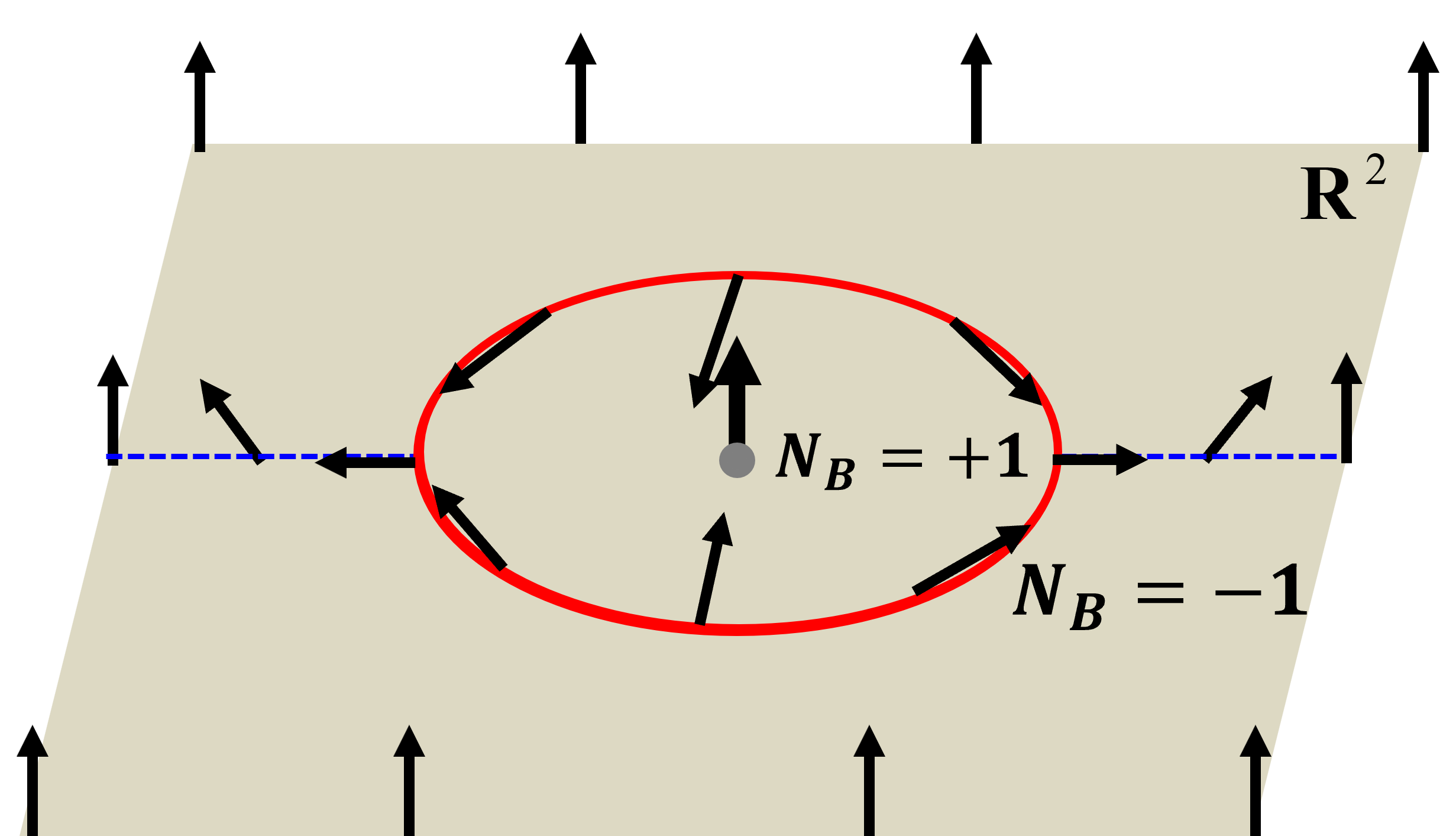}&
\includegraphics[width=6.5cm]{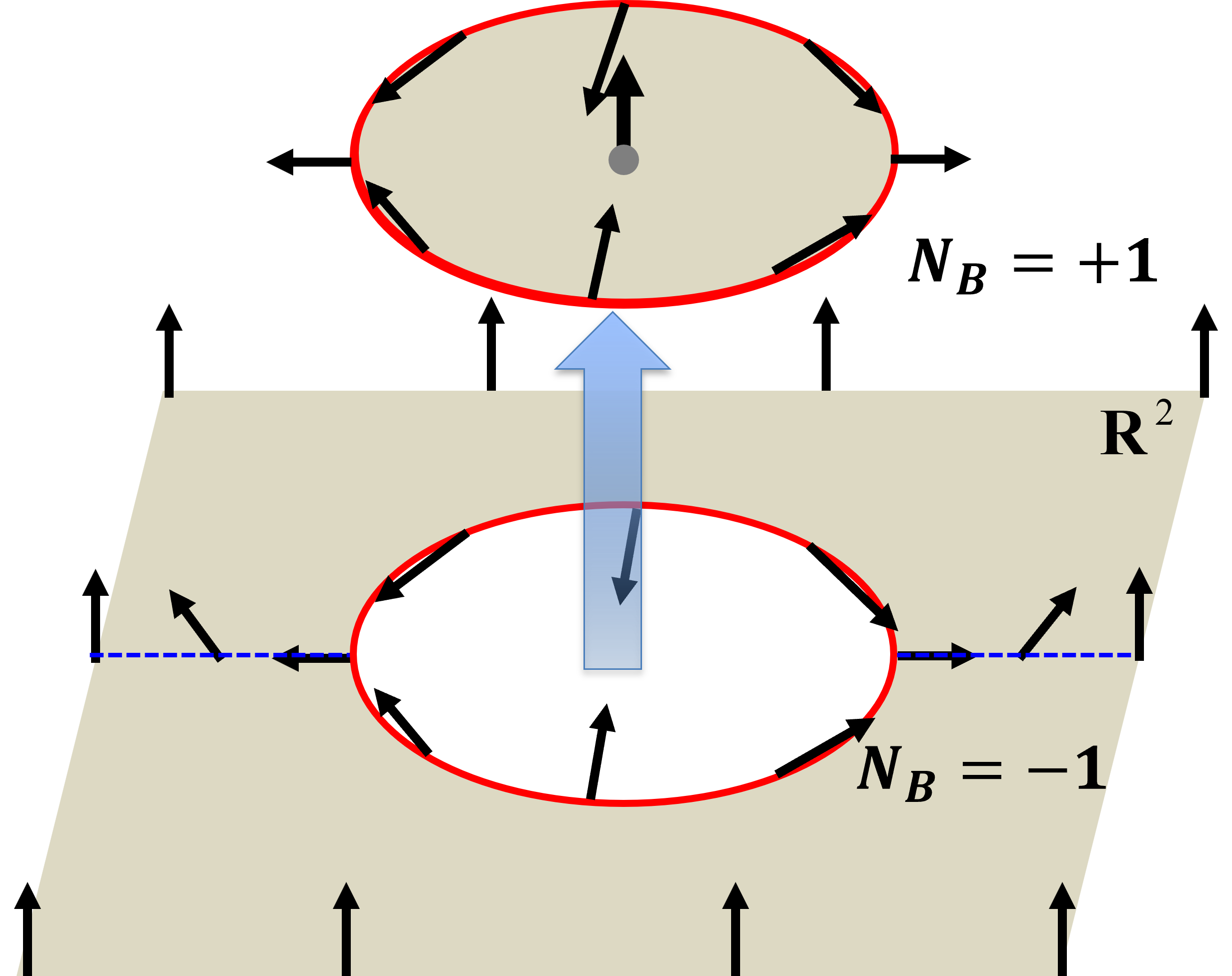}\\
(b) & (c)\\
\includegraphics[width=6.5cm]{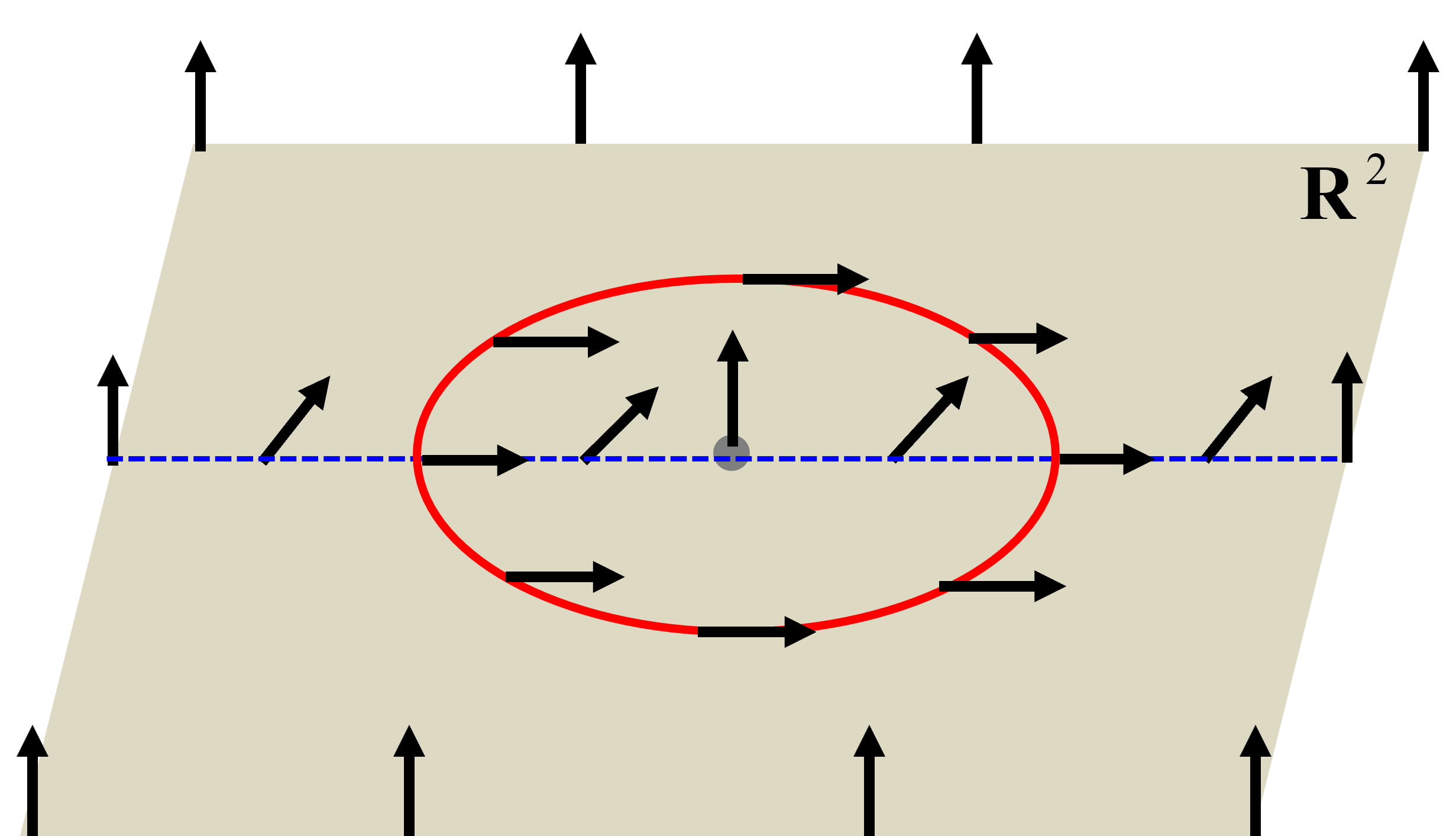}&
\includegraphics[width=6.5cm]{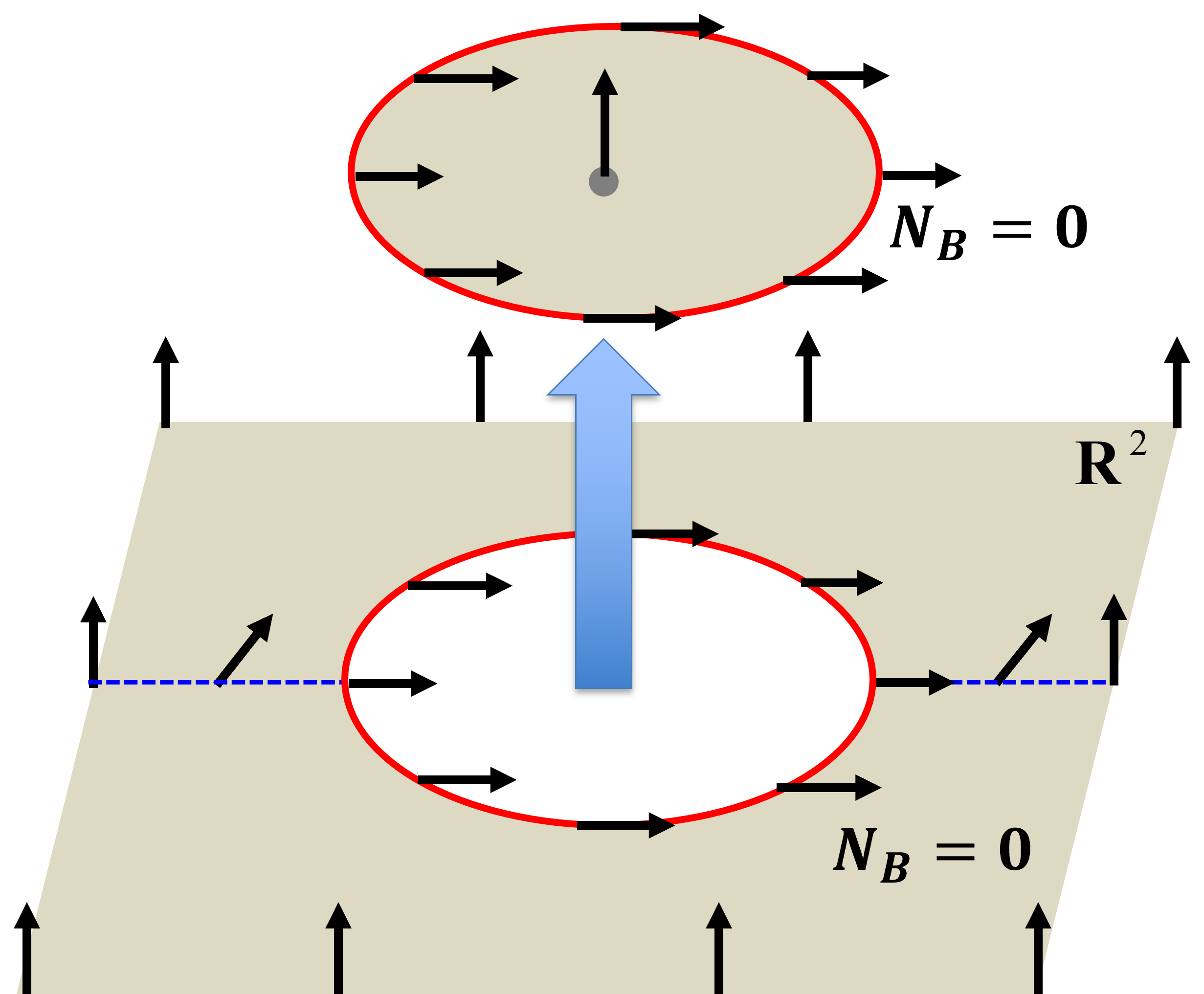}\\
(d) & (e)
\end{tabular}
\caption{Surgery of a single soliton. 
(a) A single soliton without any Skyrmion. 
(b) Before a surgery, 
one has to deform it 
to induce charged pions 
along the red circle. 
The topological lump charge of the interior 
(exterior) of the red circle is 
$k = + 1/2$ $(- 1/2)$ 
and thus the baryon number is 
$N_{\rm B} = - 1$ $(+ 1)$.
(c) Under a surgery, one can cut 
it along the red curve.
Then, the single soliton 
is decomposed into 
a pancake soliton 
of  the baryon number 
$N_{\rm B} = - 1$ 
and a soliton with a hole 
of  the baryon number 
$N_{\rm B} = + 1$. 
(d) and (e) A topologically different surgery. 
In the presence of the magnetic field in the $z$-direction, 
this process is energetically disfavored compared with 
the process in (b) and (c).
}
\label{fig:surgery2}
\end{center}
\end{figure}

\subsection{Surgery of a chiral soliton}
Once one understands the surgery of the domain-wall Skyrmions, 
then one can recognize that it is also possible to 
decompose a chiral soliton without any Skyrmions.
This can be done by exciting the charged pions 
with changing the ${\mathbb C}P^1$ moduli at 
a closed curve of a preferred surface area.
In fig.~\ref{fig:surgery2}, the surgery of a single chiral 
soliton is illustrated. 
In this case, topologically different surgeries are possible.
When the ${\mathbb C}P^1$ moduli lie along the circle, 
one can assign a winding number of the charged pions 
along the circle.
For instance, 
configurations in  fig.~\ref{fig:surgery2}~(b) and (d) have 
the winding numbers $+1$ and $0$, respectively, 
resulting in the topologically different surgeries 
in  fig.~\ref{fig:surgery2}(c) and (e), respectively.
However, energetically preferred surgery is 
the process of (b) and (c) because 
the configuration in (b) satisfies the quantization condition 
while the one in (d) cannot. 
Thus, under the surgery of a single chiral soliton, a pair creation of a baryon and anti-baryon 
occurs. 
The soliton behaves as the Dirac sea 
and the surgery provides a pair creation.

\section{Spin statistics of topological solitons}
\label{sec:spin-statistics}

In this section, we discuss the 
spin statistics of the pancake solitons, holes 
and domain-wall Skyrmions.
The statistics can be found by the responses 
of the effective action under a 
$2\pi$ rotation of the solitons.
We will show that 
the pancake solitons and holes 
with the minimal winding numbers
are fermions, 
while the domain-wall Skyrmions 
are bosons.

There are at least two ways to see the statistics of the solitons.
One is the method developed by Witten~\cite{Witten:1983tx}
(see also Ref.~\cite{Elitzur:1984kr}).
Another is to see the 
coupling of the baryon current and 
the spin structure, which 
was recently discussed by Lee, Ohmori, and Tachikawa~\cite{Lee:2020ojw}.
Here, we will discuss both methods.

\subsection{Witten's method}

In this method, we embed the configuration of $\SU(2)$-valued 
pion field to $\SU(3)$-valued one,  
and evaluate the WZW term of the fifth order of the pion field.
In this procedure, we can relate the 
spatial rotation with the rotation in 
the internal $\SU(2)$ flavor rotation.
The $2 \pi$ rotation of the solitons in the spacetime corresponds to the rotation by 
$ - \bs{1}_2$ of the flavor $\SU(2)$.
Since $\pi_4 (\SU(2)) = \bb{Z}_2$, 
the $\SU(2)$ transformation can have the 
winding in the spacetime where 
the transformation cannot be homotopically 
equivalent to the identity.
We can efficiently calculate the 
response of the effective action under 
the homotopically non-trivial $\SU(2)$
transformation using the embedding 
of $\SU(2)$ to $\SU(3)$, 
where the homotopically non-trivial $\SU(2)$
transformation can be expressed by 
homotopically trivial transformation in $\SU(3)$.
It has been found that 
the fifth order of the pion field 
in the WZW term is relevant to the response 
of the effective action.

In the following, we will calculate 
the response
in the case of the pancake solitons.
We first embed the $\SU(2)$-valued 
pion field $\Sigma \in \SU(2)$ to $U \in \SU(3)$,
\begin{equation}
 U(\bs{x}) = 
 \mtx{ \Sigma (\bs{x}) & \\ & 1},
 \label{eq:embedding}
\end{equation}
where $\Sigma (\bs{x}) $ is given by 
eqs.~\eqref{240423.1426}--\eqref{240423.1427}.
Since the spin statistics can be regarded as a response under 
$2\pi$ rotation in the space, we consider the spatial rotation 
along $\varphi $ direction with an angle $\alpha$ 
as $\varphi \to \varphi - \alpha$.
The spatial rotation can be expressed 
in terms of the flavor $\SU(2)$ transformation, 
\begin{equation}
U(\bs{x})
    \to 
     \mtx{
e^{i \alpha /2 } & 0  & 0 
\\ 
 0 &  e^{- i \alpha /2 } & 0 
\\
0 & 0 & 1}
U(\bs{x})
     \mtx{
e^{ - i \alpha /2 } & 0  & 0 
\\ 
 0 &  e^{i \alpha /2 } & 0 
\\
0 & 0 & 1},
\end{equation}
where we have used $\phi  = \nu \varphi $
with the winding number $\nu$.
We can immediately see that 
the $\alpha = 2 \pi$ rotation is given by
the transformation by $-\bs{1}_2$ in $\SU(2)$.
Therefore, the $\SU(2)$ transformation matrix
is double-valued while the transformed 
$U(\bs{x})$ is single-valued.
We can resolve the problem
using $\SU(3)$ transformation, 
\begin{equation}
     \mtx{
e^{i \alpha /2 } & 0  & 0 
\\ 
 0 &  e^{- i \alpha /2 } & 0 
\\
0 & 0 & 1}
U(\bs{x})
     \mtx{
e^{ - i \alpha /2 } & 0  & 0 
\\ 
 0 &  e^{i \alpha /2 } & 0 
\\
0 & 0 & 1}
= 
T^{-1}(\alpha) U (\bs{x}) T(\alpha)
\end{equation}
where we have introduced the single-valued 
transformation matrix,
\begin{equation}
T(\alpha)
 = 
 \mtx{
1 & 0  & 0 
\\ 
 0 & e^{i \alpha}  & 0 
\\
0 & 0 & e^{- i \alpha}} \in \SU(3)
\end{equation}
satisfying
$T(\alpha+ 2 \pi)  = T(\alpha) $.
Now, we rotate the soliton with the time coordinate 
$\alpha = t$,
and discuss the response of the effective action 
with the pancake soliton.
The non-trivial response comes from the 5-dimensional term in the WZW
action,
\begin{equation}
 \Gamma_5 = 
  \fr{i N_{\rm C}}{240\pi^2} \int d^5 x 
\epsilon^{\mu\nu\rho\sigma\tau} 
\tr (\tilde{L}_\mu \tilde{L}_\nu \tilde{L}_\rho \tilde{L}_\sigma \tilde{L}_\tau ).
\end{equation}
Here, $\tilde{L}_\mu  = \tilde{U} (x, t, \xi ) \der_\mu \tilde{U}^\dagger(x, t, \xi )$ 
is the left-invariant 1-form, 
and $\tilde{U} (x, t, \xi )$ 
is an extension of $U (\bs{x}) $ to the 5-dimensional space.
To construct $\tilde{U} (x, t, \xi )$, 
we introduce the fifth direction $\xi \in [0,1]$ along which the 
WZW term is defined.
We regard the time direction and the fifth direction as a disc, 
where the radial and 
angular coordinates are identified as 
$\xi $ and $t$, respectively.
Therefore, our four-dimensional spacetime
is defined on the boundary $\xi = 1$.
We take the configuration of $\tilde{U} (x, t, \xi )$ 
in the five-dimensional space 
as
\begin{equation}
 \tilde{U} (\bs{x},t, \xi)
  = 
V (t, \xi) 
U (\bs{x}) 
V^{-1} (t, \xi) 
\label{eq:Witten_ansatz}
\end{equation}
with 
\begin{equation}
 V (t, \xi) = 
 \mtx{
1 & 0  & 0 
\\ 
 0 & \xi e^{ - i t }  & - \sr{1- \xi^2}
\\
0 &  \sr{1- \xi^2} & \xi e^{ i t}}.
\label{eq:V}
\end{equation}
Note that $V(t, \xi)$ 
is reduced to $T^{-1}(t)$ for $\xi  = 1$.
By the extension, we have the homotopy
inequivalent $\SU(2)$ transformation in terms of the homotopy equivalent 
$\SU(3)$ transformation.
We now evaluate the WZW term.
As reviewed in appendix~\ref{WZW5}, we find that 
\begin{equation}
\begin{split}
 \Gamma_5 &
 = 
  \fr{i N_{\rm C}}{240\pi^2} \int \tr (L^5 )
+ 
\fr{ i N_{\rm C} }{24 \pi^2}
\int_{M_4} 
 \tr (U^{-1} d U)^3  \wedge \Omega)|_{\xi =1}
 \\
 &
 = 
\frac{2 \cdot 3! \cdot N_{\rm C}}{24 \pi} \int_{\bb{R}^3}
d^{3} \bs{x} 
\sin \chi \cos \chi 
\der_\rho \chi
\der_\varphi \phi
\der_z\theta .
\end{split}
\end{equation}
Note that we have used 
\begin{equation}
    \Omega |_{\xi =1} 
    = i \mtx{0 & 0 & 0 \\ 0 & -1 &  0 \\ 0 & 0 & 1 } dt 
\end{equation}
and done the integral along the time direction.
To evaluate the spatial integral, 
we substitute the configuration of the pancake soliton,
hole soliton, and domain-wall Skyrmion.
\begin{itemize}
    \item Pancake soliton. 
    The integral region along $\rho$-direction is reduced 
    to $ 0 \leq \rho \leq R$, 
\begin{equation}
\begin{split}
 \Gamma_5 &
 = 
\frac{2 \cdot 3! \cdot N_{\rm C}}{24 \pi}
\int_0^R d\rho \der_\rho \chi 
\int_{-\infty}^\infty dz \der_z \theta 
\int_0^{2 \pi} d\varphi \der_\varphi \phi 
\sin \chi \cos \chi 
= 
\pi \nu N_{\rm C}.
\end{split}
\end{equation}
Therefore, the pancake soliton with 
an odd winding number is a fermion 
while the one with an even winding number 
is a boson.
    \item Hole soliton.
In contrast to the pancake soliton, 
the integral region along $\rho$-direction is 
    to $ R \leq \rho \leq \infty$,
\begin{equation}
\begin{split}
 \Gamma_5 &
 = 
\frac{2 \cdot 3! \cdot N_{\rm C}}{24 \pi}
\int_R^\infty d\rho \der_\rho \chi 
\int_{-\infty}^\infty dz \der_z \theta 
\int_0^{2 \pi} d\varphi \der_\varphi \phi 
\sin \chi \cos \chi 
= 
  \pi \nu N_{\rm C}.
\end{split}
\end{equation}
Therefore, the hole soliton with 
an odd winding number is a fermion 
while the one with an even winding number 
is a boson.
    
    \item Domain-wall Skyrmion.
   Since the integral region along $\rho$-direction 
for the domain-wall Skyrmion is 
$ 0 \leq \rho \leq \infty$,
we have 
\begin{equation}
\begin{split}
 \Gamma_5 &
 = 
\frac{2 \cdot 3! \cdot N_{\rm C}}{24 \pi}
\int_0^\infty d\rho \der_\rho \chi 
\int_{-\infty}^\infty dz \der_z \theta 
\int_0^{2 \pi} d\varphi \der_\varphi \phi 
\sin \chi \cos \chi 
= 
 2 \pi \nu N_{\rm C}.
\end{split}
\end{equation}
Here, we have used $\chi \to \pi$ as $\rho \to \infty$.
Therefore, the domain-wall Skyrmion is 
always a boson.
\end{itemize}


\subsection{Lee-Ohmori-Tachikawa's method}

Another method to find the spin statistics 
is to see the coupling of the soliton charge 
and the spin structure.%
\footnote{The authors thank Kantaro Ohmori for discussions on this method.}
Here, the spin structure 
is roughly the 
$\bb{Z}_2$ center sector of the ${\rm Spin} (3,1)$ spin connection 1-form for the spacetime rotational symmetry.
In quantum field theories, we can understand the spin structure as 
an assignment of the periodic or anti-periodic boundary conditions for 
the fermions when we compactify one of the spacetime directions to $S^1$.
Under the $2\pi$ rotation along the $S^1$,
the fermions can receive the $\pm 1$ phase, 
and the spin structure specifies the 
phase.
Since the $\bb{Z}_2$ sector only changes 
the phase of fermions but does not change 
that of bosons, we can distinguish the spin statistics using the spin structure.

In practice, 
we can distinguish 
the spin statistics of the solitons by the response under the change of the spin structure,
i.e., the change of the boundary conditions along $S^1$ direction.
The change of the boundary condition can be described by 
a $\bb{Z}_2$ 1-form gauge field $\eta$,
\begin{eqnarray}
    \oint_{S^1} \eta \in \pi {\mathbb Z} ,
    \label{eq:eta-normalization}
\end{eqnarray}
which can be understood as an Aharonov-Bohm phase associated with the $\bb{Z}_2$ sector of 
${\rm Spin}(3,1)$ connection.

It has been recently found that 
there should be a coupling between 
the spin structure and the Skyrmion current
in the the WZW term for two flavors $N_{\rm F}=2$ in order to match the non-perturbative anomaly of the global $\SU(2)_L$ symmetry~\cite{Lee:2020ojw}.
The explicit form of the 
effective action is 
\begin{equation}
       e^{ i S^{(2)}_{\rm WZW} [ \frak{s}]}
       = (-1)^{N_{\rm C} [\Sigma : M_4 \to {\rm SU}(2)]},
\end{equation}
where $[\Sigma : M_4 \to {\rm SU}(2)]$ is an equivalence class 
specified by the reduced bordism group $\tilde{\Omega}^{\rm spin}_4 ({\rm SU}(2)) = \bb{Z}_2$.
While the effective action may be abstract, 
we can express the effective action in the point particle limit of the Skyrmion using the holonomy associated with the spin connection of the spacetime.
We introduce a holonomy of ${\rm Spin}(3,1)$ gauge field, 
\begin{equation}
{\rm hol}(\frak{s}, {\cal C})    := 
{\rm P}\exp \left( i \int_{\cal C} \fr{1}{2!}\omega_{\mu} {}^{ab} \Sigma_{ab} dx^\mu \right).
\end{equation}
Here, 
$\frak{s}$  in the left-hand side specifies the 
configuration of the link variable given by the spin connection,
$\omega_{\mu} {}^{ab}$ 
is the spin connection of the spacetime given by 
the vielbein $e_{\mu }{}^a$
with the local Lorentz indices $a, b ,... = 0,1,2,3$, 
the symbol $\Sigma_{ab}$ is the generator of ${\rm Spin}(3,1) $,
${\cal C}$ denotes a closed worldline of the Skyrmion, 
and 
``${\rm P}$'' represents the path ordered product.
In the flat spacetime, the holonomy can be reduced to 
$\bb{Z}_2 \subset {\rm Spin}(3,1) $ rather than the identity,
\begin{equation}
{\rm hol}(\frak{s}, {\cal C})    := 
{\rm P}\exp \left( i \int_{\cal C} \fr{1}{2!}\omega_{\mu} {}^{ab} \Sigma_{ab} dx^\mu \right)
= (-1)^{\epsilon ({\cal C}) } \bs{1}_{{\rm Spin}(3,1) },
\end{equation}
with $\epsilon ({\cal C})= 0,1$.
Now, we can have a WZW 
action for the Skyrmion 
using the holonomy on ${\cal C}$
associated with ${\rm Spin} (3,1)$ as 
\begin{equation}
       e^{ i S^{(2)}_{\rm WZW} [ \frak{s}]}
       = (-1)^{ N_{\rm B} N_{\rm C} \epsilon ({\cal C}) } .
\end{equation}
The multiplication by $N_{\rm B}$
comes from the baryon number specified by ${\rm SU}(2)$, 
and that by $N_{\rm C}$ comes from the 
number of quarks, which is the same as the conventional WZW term.
This effective action means that the Skyrmion 
has a spin, and its statistics can be classified by the color $N_{\rm C}$ as well as the baryon number.

We can see the spin statistics of the Skyrmion 
by the change of the configuration of the 
link variable $\frak{s}$.
By changing $\epsilon({\cal C})$
by 
Aharonov-Bohm phase of ${\rm Spin}(3,1)$, 
the effective action is changed as 
\begin{equation}
   e^{ i S^{(2)}_{\rm WZW} [ \frak{s}]}
\to 
e^{ i N_{\rm B} N_{\rm C} \int_{\cal C} \eta} 
   e^{ i S^{(2)}_{\rm WZW} [ \frak{s}]}
   .
\end{equation}
Since the response $e^{ i N_{\rm B}N_{\rm C} \int_{\cal C} \eta} $ 
is Abelian, we can now recover the 
configuration of the Skyrmion from 
the point-like one to a general one as
\begin{equation}
   e^{ i S^{(2)}_{\rm WZW} [ \frak{s}]}
\to 
\exp\left( i N_{\rm C} \int_{M_4} \eta \wedge
\fr{1}{4\pi^2 \cdot 3!} \tr [(\Sigma d\Sigma^\dagger )^3] \right)
   e^{ i S^{(2)}_{\rm WZW} [ \frak{s}]}
   .
    \label{eq:WZW-2}
\end{equation}

This response clearly shows that 
the Skyrmions with the odd Skyrmion number 
are fermions when $N_{\rm C}$ is odd.
To see this, 
we compactify the time direction to 
$S^1$, i.e., 
$M_4 = S^1 \times \bb{R}^3$,
take $\eta$ as
\begin{equation}
    \int_{S^1} \eta= \pi ,
\end{equation}
and consider the response of the effective action,
\begin{equation}
\begin{split}
&
\exp\left( i N_{\rm C} \int_{M_4} \eta \wedge
\fr{1}{4\pi^2 \cdot 3!} \tr [(\Sigma d\Sigma^\dagger )^3] \right)
             \\
             & =
             \exp 
             \left(
             i N_{\rm C}
             \int_{S^1} \eta
             \cdot 
             \int_{M_3} 
             {\rm tr} [(\Sigma d\Sigma^\dagger )^3]\right)
           \\
           &
           = 
              \exp 
             \left(
             i \pi N_{\rm C}
\int_{M_3} d^3\bs{x} \sin \chi \cos \chi 
\der_\rho \chi
\der_\varphi \phi
\der_z\theta 
\right).
\end{split}
\end{equation}
By substituting the 
configuration of 
the pancake soliton, 
hole soliton, and 
domain-wall Skyrmion, 
we find that 
the pancake soliton 
and the hole soliton with 
odd winding numbers are 
fermions, 
and the others are bosons.

One comment is in order.
From the effective action
with the spin structure, 
we have the coupling between 
the spin structure and 
the topological charge 
by integrating the direction 
along the domain wall is localized.
To this end, let us calculate a contribution from eq.~(\ref{eq:WZW-2}) to the soliton effective Lagrangian 
by the integration of 
eq.~(\ref{eq:WZW-2}) 
over the codimension $x^3$ of the soliton:
\begin{eqnarray}
\int dx^3 {\cal L}_{\rm WZW}^{(2)}
=  N_{\rm C} 
\eta^0 
\int dx^3 
    {\cal B}
= 2 N_{\rm C}  q \eta^0
\end{eqnarray}
with the topological 
lump charge $q$ in eq.~(\ref{eq:pi2}).
Equivalently, it is in the effective action
\begin{eqnarray}
 S_{\rm WZW}^{(2)} = 2 N_{\rm C} \int_{M_3}\eta \wedge {\cal Q} 
,
\end{eqnarray}
with a pullback of the standard area form on $S^2$, 
\begin{equation}
    {\cal Q} \equiv \frac{1}{8\pi} \bm{n}\cdot \left(\partial_i \bm{n}\times \partial_j \bm{n} \right)dx^i \wedge dx^j
    =\frac{1}{8\pi} \bm{n}\cdot \left(d \bm{n}\times d \bm{n} \right) \,.
\end{equation}
The Aharonov-Bohm phase factor is  
\begin{eqnarray}
  &&  \exp \left( - i N_{\rm C} \int dt \int d^2x \, \int dx^3 {\cal L}_{\rm WZW}^{(2)} \right)
 = \exp \left( - i N_{\rm C} \int dt \, \eta^0 \int d^2x \, 2 q \right)
 \nonumber\\
&&  = \exp \left( - 2 \pi i N_{\rm C}   k \right) = 1 . 
\end{eqnarray}
Therefore, a single lump $(k=1)$ (thus any number $k$ of lumps) is a {\it boson} for any 
$N_{\rm C}$. 
This is as expected because 
the single lump in the soliton is a double baryon bound state of 
$N_{\rm B}=2$, 
and more generally even quantization 
in eq.~(\ref{eq:NB}).

\subsection{Comments on surgeries}

Let us make comments on surgeries 
discussed in the last section. 
In fig.~\ref{fig:surgery1}, 
the domain-wall Skyrmion has 
the baryon number $N_{\rm B}=2$ and is a boson,
while the hole and pancake soliton have 
the baryon numbers $N_{\rm B}=1$ and are fermions.

\begin{figure}[t]
\begin{center}
\includegraphics[width=16cm]{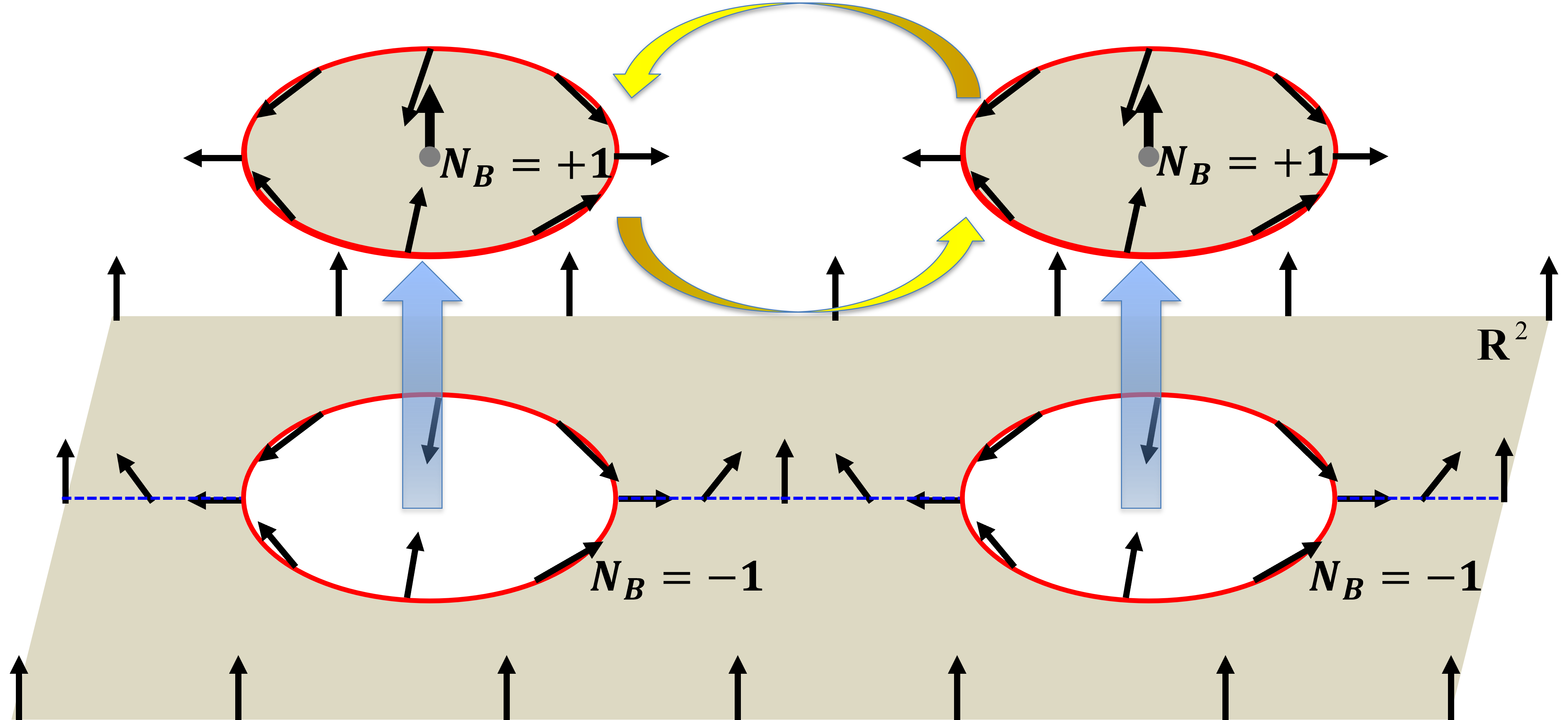} 
\caption{
Exchange of two parts of a single chiral soliton. 
The procedure is: 
(1) cut the two parts by surgeries,  
(2) exchange them leaving two holes, 
(3) glue them to the holes.
This procedure yields a minus sign for exchanging two fermions. 
\label{fig:exchange}
}
\end{center}
\end{figure}

One may consider the exchange 
of two parts of the chiral soliton 
(without domain-wall Skyrmions).
First, cut the two parts by surgeries 
as in fig.~\ref{fig:surgery2},  
exchange them leaving two holes, 
and glue them to the holes 
as illustrated in fig.~\ref{fig:exchange}.
We then obtain a minus sign for exchanging two fermions. 
Therefore, 
each part of the chiral soliton 
is a fermion.

\section{Summary and Discussion}
\label{sec:summary}

In this paper, we have clarified the spin statistics of 
chiral solitons of finite sizes (pancake solitons),  
holes on a chiral soliton, and
domain-wall Skyrmions
in QCD with $N_{\rm F}=2$ flavors 
at finite baryon density $\mu_{\rm B}$
in magnetic field $B$. 
In order to overcome the absence of the 
WZW term for two flavors,
we first have used the conventional Witten's method to embed $\SU(2)$ chiral Lagrangian to 
the $\SU(3)$ one with the WZW term. 
Then, we have clarified spin statistics 
of topological solitons: 
domain-wall Skyrmions are bosons 
and pancake solitons or holes 
of the size of $S_0=2\pi/e B_z$ are fermions 
with baryon number $N_{\rm B}=1$.
Larger pancake solitons or holes 
of the area of odd (even) multiple of $S_0=2\pi/e B_z$ 
are fermions (bosons).
We also have used 
the $N_{\rm F} = 2$ WZW term \cite{Lee:2020ojw} 
constructed in terms of 
spin structures and obtained the same results.
We also have constructed the low-energy effective theory 
on a single chiral soliton,  
obtained 
the ${\mathbb C}P^1$ model 
with topological terms originated from the WZW term, 
to confirm that 
domain-wall Skyrmions are bosons. 
We also have proposed surgeries of topological solitons.
Domain-wall Skyrmions with $N_{\rm B}=2$ (bosons) can be decomposed into 
a chiral soliton with a topological hole  
with $N_{\rm B}=1$ (fermion) 
and a pancake soliton 
with $N_{\rm B}=1$ (fermion) 
as in fig.~\ref{fig:surgery1}, 
while 
a chiral soliton without any Skyrmion 
can be decomposed into 
a chiral soliton with a hole with 
$N_{\rm B}=\pm 1$ (fermion) 
and a pancake soliton with $N_{\rm B}=\mp 1$ (fermion) as in fig.~\ref{fig:surgery2}.
The surgery allows the exchange of 
two quantized areas of a single chiral soliton, showing a fermion statistics, as illustrated in fig.~\ref{fig:exchange}.

Here, we address several discussions 
and possible future directions.
One of the possible future directions is 
the extension to the three flavor case, $N_{\rm F} =3$.
In this case, the ${\mathbb C}P^2$ moduli appear on a chiral soliton.

As for surgeries of topological solitons, 
only topological aspects and spin statistics 
have been studied in this paper. 
Dynamically there should exist an attractive interaction between 
separated components 
implying the energy barrier
that prevents surgeries and ensuring the stability of chiral solitons and domain-wall Skyrmions.
Calculation of the interaction is one of the important future directions,
and one also should be able to calculate 
quantum tunneling rates 
\cite{Eto:2022lhu}.

Aside from spin statistics of topological solitons, 
this paper gives a new aspect of the ${\mathbb C}P^1$ model: 
the ${\mathbb C}P^1$ model on two dimensional plane 
with a boundary 
as in fig.~\ref{fig:pion-string}, 
or on a disk as in fig.~\ref{fig:pancake}. 
As a similar situation, the ${\mathbb C}P^{N-1}$ model on a finite interval 
was discussed in the literature
\cite{Milekhin:2012ca,Monin:2015xwa,Bolognesi:2016zjp,Milekhin:2016fai,Betti:2017zcm,Flachi:2017xat,Bolognesi:2018njt,Yoshii:2019yln,Flachi:2019jus}. 
In particular, 
the ${\mathbb C}P^{N-1}$ model 
on a disk was studied in Ref.~\cite{Pikalov:2017lrb}.
However, as far as we know,  
the case with a U(1) degrees of freedom remaining on the boundary 
as in figs.~\ref{fig:pion-string} and \ref{fig:pancake}
has not been 
considered yet.

Before closing this paper, let us 
comment on a domain-wall Skyrmion in QCD matter under rapid rotation~\cite{Eto:2023tuu}. 
The anomalous term for the $\eta'$ meson  \cite{Huang:2017pqe,Nishimura:2020odq} 
reproducing the chiral vortical effect~\cite{Vilenkin:1979ui,Vilenkin:1980zv,Son:2009tf,Landsteiner:2011cp,Landsteiner:2012kd,Landsteiner:2016led} at low energy  
leads to a CSL composed of the $\eta'$ meson 
(or $\eta$ meson for the two-flavor case) 
under rapid rotation,  
instead of CSL of the neutral pion in the magnetic field ~\cite{Huang:2017pqe,Nishimura:2020odq,Chen:2021aiq}. 
In a large parameter region
in the case of two flavors, 
 a single $\eta$-soliton energetically decays into a pair of non-Abelian solitons by a neutral pion condensation 
 \cite{Eto:2021gyy}.
  The vector symmetry ${\rm SU}(2)_{\rm V}$ is spontaneously broken down to its U$(1)$ subgroup 
  in the vicinity of each non-Abelian soliton,
  resulting in ${\mathbb C}P^1$ NG modes localized around it \cite{Eto:2021gyy,Eto:2023rzd}, 
 and yielding ${\mathbb C}P^1$ moduli~\cite{Nitta:2014rxa,Eto:2015uqa,Nitta:2015mma,Nitta:2015mxa,Nitta:2022ahj}. With the single-soliton approximation, the domain-wall Skyrmion  phase for rapid rotations was found in ref.~\cite{Eto:2023tuu}. 
  Unlike the case of the magnetic field studied in this paper, one lump on the soliton corresponds to one Skyrmion in the bulk in this case. 
Therefore, we expect that 
a domain-wall Skyrmion must be a fermion 
in this case, 
in contrast to bosons for magnetic field 
uncovered in this paper.

In the case of a single flavor $N_{\rm F}=1$, 
the degree of the $\eta$ meson can be used for constructing baryons as 
the $\eta$ soliton \cite{Komargodski:2018odf,Karasik:2020pwu}. In such a case, Skyrmions are realized  
as vortices on the $\eta$ soliton,  
and the anyonic nature of quarks was shown \cite{Lin:2023qya}. 
Unified understanding including this case would be one of the important future directions.

\begin{acknowledgments}
We thank Minoru Eto 
and Zebin Qiu for useful discussion, 
and especially 
Kantaro Ohmori for 
useful discussion on
the $N_{\rm F}=2$ WZW term and 
spin structure.
This work is supported in part by 
 JSPS KAKENHI [Grants No.~JP23KJ1881 (YA), 
 JP22H01221 (MN), 
 and JP21K13928 (RY)] and the WPI program ``Sustainability with Knotted Chiral Meta Matter (SKCM$^2$)'' at Hiroshima University (MN).
\end{acknowledgments}

\appendix
\section{Calculation of the spin statics of pancake soliton\label{WZW5}}

In this appendix, we examine the statistics of pancake soliton by evaluating rather the WZW action that Witten used for the discussion of the statistics of Skyrmions \cite{Witten:1983tx,Witten:1983tw} than the $N_\mathrm{F}=2$ WZW term~\eqref{eq:WZW-2}.
For this purpose, we apply the method described in Ref.~\cite{Balachandran:1985fb}.
As Witten did, we extend the space-time $M_4 = S^1\times M_3$ to $M_5 = D_2 \times M_3$, where $M_3$ is the physical space and $D_2$ stands for a two-dimensional disc parameterized by (normalized) time $t\in[0,2\pi]$ as an axis variable and $\xi\in[0,1]$ as a radial variable. 
We consider an $\SU(3)$ chiral field on the five dimensional space-time $M_5$ defined in eq.~\eqref{eq:Witten_ansatz}, i.e.,  $
\tilde{U}(\bm{x}, t, \xi)  =V(t, \xi) U(\bm{x}) V^{\dagger}(t, \xi) $,
where $U(\bm{x})$ stands for a static Skyrmion field and $V$ is a collective coordinate defined on $D_2$. The variable $V$ is assumed to satisfy the boundary condition $V(0,\xi)=V(2\pi,\xi)$.
The left invariant Maurer-Cartan form of $\tilde{U}$ can be written as
\begin{align}
\tilde{L} & =\tilde{U} d \tilde{U}^\dagger =V\left(U \Omega U^\dagger+L-\Omega\right) V^\dagger
\end{align}
where $L=U d U^\dagger,$ and $\Omega=V^\dagger d V$.
Using this, we obtain
\begin{equation}
\begin{aligned}
\operatorname{Tr}\left(\tilde{L}^{5}\right)= & \operatorname{Tr}\left(L^{5}\right) \\
& +5 d \operatorname{Tr}\left[L^{3} \Omega-\frac{1}{2}(L \Omega)^{2}+L \Omega^{3}\right. \\
& \qquad\qquad\quad+U(L-\Omega)^{3} U^\dagger \Omega  \left.+\frac{1}{2}\left(U(L-\Omega) U^\dagger \Omega\right)^{2}+U(L-\Omega) U^{\dagger} \Omega^{3}\right]
\end{aligned}
\end{equation}
where we also employed 
\begin{align}
& d L=-L^{2} , \qquad
d \Omega=-\Omega^{2} ,\qquad 
 d U =-LU, \qquad
 d U^\dagger=U^\dagger L .
\end{align}
It implies that the WZW term can be written as
\begin{align}
 & \Gamma_5 \equiv \frac{iN_{\rm C}}{240 \pi^{2}} \int_{M_{5}} \operatorname{Tr}\left(\tilde{L}^{5}\right) 
 \notag\\
& =\frac{iN_{\rm C}}{240 \pi^{2}} \int_{M_{5}} \operatorname{Tr}\left(L^{5}\right) 
\notag\\
& \quad +\frac{iN_{\rm C}}{48 \pi^{2}} \int_{M_{4}} \operatorname{Tr}\left[L^{3} \Omega-\frac{1}{2}(L \Omega)^{2}+L \Omega^{3}\right. \\
& \qquad\qquad\qquad\qquad \left.
+U^\dagger(L-\Omega)^{3} U \Omega \left.+\frac{1}{2}\left(U^\dagger(L-\Omega) U \Omega\right)^{2}+U^\dagger(L-\Omega) U \Omega^{3}\right]\right|_{\xi=1}. 
\notag
\end{align}
Taking space-time dependency of $L$ and $\Omega$ into account, i.e., $L^{4} =\Omega^{2}|_{\xi=1}=0
$, this can be simplified as
\begin{align}
\Gamma_5
& =\frac{iN_{\rm C}}{48 \pi^{2}} \int_{M_{4}} \operatorname{Tr}\left(L^{3} \Omega+U^\dagger L^{3} U \Omega\right) 
=\frac{iN_{\rm C}}{48 \pi^{2}} \int_{M_{4}} \operatorname{Tr}\left[(L^{3} +R^{3} )\Omega\right]
\end{align}
where $R=d U^\dagger U$.
Expanding $\Omega_0$ in terms of the Gell-Mann matrices $\lambda_\alpha$ ($\alpha=1,2,...,8$) as
\begin{equation}
\Omega_{0}=V^\dagger \partial_{0} V=\frac{\lambda_{\alpha}}{2} \operatorname{Tr}\left(\lambda_{\alpha} V^\dagger \partial_{0} V\right)=\frac{i}{2} \lambda_{\alpha} \omega^{\alpha} \ ,
\end{equation}
we obtain
$$
\Gamma_5
=+\frac{N_{\rm C}}{96 \pi^{2}} \int d^{4} x ~ \varepsilon^{i j k} \operatorname{Tr}\left[\left(L_{i} L_{j} L_{k}+R_{i} R_{j} R_{k}\right) \lambda_\alpha\right] \omega^{\alpha}
$$
where we used $dx^\mu\wedge dx^\nu\wedge dx^\alpha \wedge dx^\beta=dx^4 \varepsilon^{\mu\nu\alpha\beta}$.
In order to evaluate the WZW term for $\SU(2)$ solitons, we consider the $\SU(3)$ chiral field into which the $\SU(2)$ chiral field $\Sigma$ is trivially embedded as defined in eq.~\eqref{eq:embedding}.
Then, the Maurer-Cartan forms can simply be written as
$$
L=\left(\begin{array}{cc}
\Sigma d \Sigma^\dagger & 0 \\
0 & 0
\end{array}\right), \qquad
R=\left(\begin{array}{cc}
d \Sigma^\dagger\Sigma & 0 \\
0 & 0
\end{array}\right) \ .
$$
We decompose their elements in terms of the Pauli matrices $\tau_a$ as
\begin{equation}
\begin{aligned}
& \Sigma \partial_{i} \Sigma^\dagger=\frac{\tau_{a}}{2} \operatorname{Tr}\left(\tau_{a} \Sigma \partial_{i} \Sigma^\dagger\right)\equiv\frac{i}{2} \tau_{a} \xi_{i}^{a} \ , \\
& \partial_{i} \Sigma^\dagger \Sigma=\frac{\tau_{a}}{2} \operatorname{Tr}\left(\tau_{a} \partial_{i} \Sigma^\dagger \Sigma\right)\equiv\frac{i}{2} \tau_{a} \zeta_{i}^{a} \ .
\end{aligned}
\end{equation}
Using the decomposition, one obtains after simple algebra that
\begin{equation}
\begin{aligned}
& \varepsilon^{i j k} \Sigma \partial_{i} \Sigma^\dagger \Sigma \partial_{j} \Sigma^\dagger \Sigma \partial_{k} \Sigma^\dagger
 =\frac{1}{8} \varepsilon^{i j k} \varepsilon_{a b c} \xi_{i}^{a} \xi_{j}^{b} \xi_{k}^{c} \mathbf{1}_2 \ ,
\\
&\varepsilon^{i j k} \partial_{i} \Sigma^\dagger \Sigma \partial_{i} \Sigma^\dagger \Sigma \partial_{k} \Sigma^\dagger \Sigma =\frac{1}{8} \varepsilon^{i j k} \varepsilon_{a b c} \zeta_{i}^{a} \zeta_{j}^{b} \zeta_{k}^{c} \mathbf{1}_2 \ ,
\end{aligned}
\end{equation}
where $\mathbf{1}_2$ is the $2\times2$ unit matrix.
It indicates that the baryon number $N_{\rm B}\in \pi_3 [\SU(2)]$ defined in eq.~\eqref{eq:def_baryon_number} can be cast into the form
\begin{align}
N_{\rm B} & 
 =\frac{1}{96 \pi^{2}} \int d^{3} x ~\varepsilon^{i j k} \varepsilon_{a b c} \xi_{i}^{a} \xi_{j}^{b} \xi_{k}^{c} 
 =\frac{1}{96 \pi^{2}} \int d^{3} x ~\varepsilon^{i j k} \varepsilon_{a b c} \zeta_{i}^{a} \zeta_{j}^{b} \zeta_{k}^{c}.
\end{align}
Consequently, we find that the WZW term can be written as
\begin{align}
\Gamma_5
& =\frac{N_{\rm C}}{96 \pi^{2}} \int d^{4} x ~ \frac{1}{8} \varepsilon^{i j k} \varepsilon_{a b c}\left(\xi_{i}^{a} \xi_{j}^{b} \xi_{k}^{c}+\zeta_{i}^{a} \zeta_{j}^{b} \zeta_{k}^{c}\right) \operatorname{Tr}\left(h \lambda_{\alpha}\right) \omega^{\alpha} 
\notag\\
& =\frac{N_{\rm C}N_{\rm B}}{4}  \int d x^{0} \operatorname{Tr}\left(h \lambda_{\alpha}\right) \omega^{\alpha} 
\notag\\
& =\frac{N_{\rm C}N_{\rm B}}{2 \sqrt{3}} \int d x^{0} \omega^{8}
\end{align}
where $h=\operatorname{diag}(1,1,0)$.

Let us evaluate the WZW term using the explicit form of $V$ given in eq.~\eqref{eq:V}. 
Since $\Omega_0$ is given by
\begin{equation}
    \Omega_0 |_{\xi=1}=V^\dagger\partial_t V|_{\xi=1} =\operatorname{diag}(0,-i,i) \ ,
\end{equation}
we obtain
\begin{equation}
    \omega^8=-i\operatorname{Tr}(\lambda_8\Omega_0)|_{\xi=1}=-\sqrt{3}.
\end{equation}
Therefore, we find that the WZW term takes
\begin{equation}
    S_{\text{WZW}} =-\pi N_{\rm C}N_{\rm B} \ .
\end{equation}

Let us explicitly calculate the WZW term of the pancake solitons. With the parameterization of the $\SU(2)$ chiral field \eqref{eq:parametrization_Sigma_pancake}, the $\SU(3)$ chiral field for pancake solitons is given by
\begin{equation}
 U (\bs{x})
 = 
\mtx{
\sigma + i\pi^0 & i(\pi^1 - i \pi^2) & 0 
\\ 
i(\pi^1 + i \pi^2)  & \sigma - i\pi^0  & 0 
\\
0 & 0 & 1}
 = 
\mtx{
e^{i\theta }  \cos\chi 
 & i e^{- i  \phi }   \sin \chi & 0 
\\ 
i e^{i \phi } \sin \chi   & e^{ - i\theta }  \cos\chi   & 0 
\\
0 & 0 & 1}
\end{equation}
Here, $\chi= \chi (\rho)$, $\theta = \theta (z)  $,
and $\phi = \phi (\varphi) = \nu \varphi$.
By explicit calculation, one gets
\begin{equation}
\begin{split}
 U \der_\rho U^\dagger
&
=
\mtx{
e^{i\theta }  \cos\chi 
 & i e^{- i  \phi }   \sin \chi & 0 
\\ 
i e^{i \phi } \sin \chi   & e^{ - i\theta }  \cos\chi   & 0 
\\
0 & 0 & 1}
\der_\rho 
\mtx{
e^{ - i\theta }  \cos\chi 
 &  - i e^{- i  \phi }   \sin \chi & 0 
\\ 
 - i e^{i \phi } \sin \chi   & e^{ i\theta }  \cos\chi   & 0 
\\
0 & 0 & 1}
\\
&
=
\mtx{
e^{i\theta }  \cos\chi 
 & i e^{- i  \phi }   \sin \chi & 0 
\\ 
i e^{i \phi } \sin \chi   & e^{ - i\theta }  \cos\chi   & 0 
\\
0 & 0 & 1}
\mtx{
 - e^{ - i\theta }  \sin \chi 
 &  - i e^{- i  \phi }   \cos \chi & 0 
\\ 
 - i e^{i \phi } \cos \chi   & 
 - e^{ i\theta }  \sin\chi   & 0 
\\
0 & 0 & 0}
\der_\rho \chi
\\
&
=
\mtx{
0 
 &   -i e^{i  (\theta - \phi ) }  & 0 
\\ 
 -i e^{ - i  (\theta - \phi ) }  & 0   & 0 
\\
0 & 0 & 0}
\der_\rho \chi \,,
\end{split}
\end{equation}
\begin{equation}
\begin{split}
 U \der_\varphi U^\dagger
&
=
\mtx{
e^{i\theta }  \cos\chi 
 & i e^{- i  \phi }   \sin \chi & 0 
\\ 
i e^{i \phi } \sin \chi   & e^{ - i\theta }  \cos\chi   & 0 
\\
0 & 0 & 1}
\der_\varphi 
\mtx{
e^{ - i\theta }  \cos\chi 
 &  - i e^{- i  \phi }   \sin \chi & 0 
\\ 
 - i e^{i \phi } \sin \chi   & e^{ i\theta }  \cos\chi   & 0 
\\
0 & 0 & 1}
\\
&
=
\mtx{
e^{i\theta }  \cos\chi 
 & i e^{- i  \phi }   \sin \chi & 0 
\\ 
i e^{i \phi } \sin \chi   & e^{ - i\theta }  \cos\chi   & 0 
\\
0 & 0 & 1}
\mtx{
0
 &  - e^{- i  \phi }   \sin \chi & 0 
\\ 
  e^{i \phi } \sin \chi   &  0
  & 0 
\\
0 & 0 & 0}
\der_\varphi \phi
\\
&
=
\mtx{
i \sin^2 \chi 
 &   - e^{ i  (\theta - \phi ) } \sin \chi   \cos \chi  & 0 
\\ 
e^{ - i  (\theta - \phi ) }  \sin \chi   \cos \chi  &
 -i \sin^2 \chi  & 0 
\\
0 & 0 & 0}
\der_\varphi \phi \,,
\end{split}
\end{equation}
\begin{equation}
\begin{split}
 U\der_z U^\dagger
&
=
\mtx{
e^{i\theta }  \cos\chi 
 & i e^{- i  \phi }   \sin \chi & 0 
\\ 
i e^{i \phi } \sin \chi   & e^{ - i\theta }  \cos\chi   & 0 
\\
0 & 0 & 1}
\der_z 
\mtx{
e^{ - i\theta }  \cos\chi 
 &  - i e^{- i  \phi }   \sin \chi & 0 
\\ 
 - i e^{i \phi } \sin \chi   & e^{ i\theta }  \cos\chi   & 0 
\\
0 & 0 & 1}
\\
&
=
\mtx{
e^{i\theta }  \cos\chi 
 & i e^{- i  \phi }   \sin \chi & 0 
\\ 
i e^{i \phi } \sin \chi   & e^{ - i\theta }  \cos\chi   & 0 
\\
0 & 0 & 1}
\mtx{
 -i e^{ - i\theta }  \cos\chi 
 &  0  & 0 
\\ 
0   &  i e^{ i\theta }  \cos\chi   & 0 
\\
0 & 0 & 0}
\der_z \theta
\\
&
=
\mtx{
 - i\cos^2 \chi 
 & -  e^{ i  (\theta  - \phi) }  \sin\chi  \cos \chi  & 0 
\\ 
 e^{-  i  (\theta  - \phi) }  \sin\chi  \cos \chi  &
   i\cos^2 \chi   & 0 
\\
0 & 0 & 0}
\der_z \theta \,,
\end{split}
\end{equation}
and 
\begin{equation}
\begin{split}
&
 U \der_\rho U^\dagger
 U \der_\varphi U^\dagger
 U \der_z U^\dagger
\\
&
= 
\mtx{
0 
 &   -i e^{i  (\theta - \phi ) }  & 0 
\\ 
 -i e^{ - i  (\theta - \phi ) }  & 0   & 0 
\\
0 & 0 & 0}
\mtx{
i \sin^2 \chi 
 &   - e^{ i  (\theta - \phi ) } \sin \chi   \cos \chi  & 0 
\\ 
e^{ - i  (\theta - \phi ) }  \sin \chi   \cos \chi  &
 -i \sin^2 \chi  & 0 
\\
0 & 0 & 0}
\\
&
\quad
\times
\mtx{
 - i\cos^2 \chi 
 & -  e^{ i  (\theta  - \phi) }  \sin\chi  \cos \chi  & 0 
\\ 
 e^{-  i  (\theta  - \phi) }  \sin\chi  \cos \chi  &
   i\cos^2 \chi   & 0 
\\
0 & 0 & 0}
\der_\rho \chi
\der_\varphi \phi
\der_z \theta
\\
&
= 
\mtx{
- i \sin \chi \cos \chi
 &   -  e^{i  (\theta - \phi) } \sin^2 \chi & 0 
\\ 
 e^{ - i  (\theta  - \phi) } \sin^2 \chi  &  i \sin \chi \cos \chi  & 0 
\\
0 & 0 & 0}
\\
&
\quad
\times
\mtx{
 - i\cos^2 \chi 
 & -  e^{ i  (\theta  - \phi) }  \sin\chi  \cos \chi  & 0 
\\ 
 e^{-  i  (\theta  - \phi) }  \sin\chi  \cos \chi  &
   i\cos^2 \chi   & 0 
\\
0 & 0 & 0}
\der_\rho \chi
\der_\varphi \phi
\der_z\theta 
\\
&
= 
-  \sin \chi \cos \chi 
\der_\rho \chi
\der_\varphi \phi
\der_z\theta 
\mtx{ 1 && \\ & 1 & \\ &&0},
\end{split}
\end{equation}
\begin{equation}
\begin{split}
&
 U \der_\varphi U^\dagger
 U \der_\rho U^\dagger
 U \der_z U^\dagger
\\
&
= 
\mtx{
i \sin^2 \chi 
 &   - e^{ i  (\theta - \phi ) } \sin \chi   \cos \chi  & 0 
\\ 
e^{ - i  (\theta - \phi ) }  \sin \chi   \cos \chi  &
 -i \sin^2 \chi  & 0 
\\
0 & 0 & 0}
\mtx{
0 
 &   -i e^{i  (\theta - \phi ) }  & 0 
\\ 
 -i e^{ - i  (\theta - \phi ) }  & 0   & 0 
\\
0 & 0 & 0}
\\
&
\quad
\times
\mtx{
 - i\cos^2 \chi 
 & -  e^{ i  (\theta  - \phi) }  \sin\chi  \cos \chi  & 0 
\\ 
 e^{-  i  (\theta  - \phi) }  \sin\chi  \cos \chi  &
   i\cos^2 \chi   & 0 
\\
0 & 0 & 0}
\der_\rho \chi
\der_\varphi \phi
\der_z \theta
\\
&
= 
\mtx{
i \sin \chi \cos \chi
 &    e^{i  (\theta - \phi) } \sin^2 \chi & 0 
\\ 
-  e^{ - i  (\theta  - \phi) } \sin^2 \chi  & 
- i \sin \chi \cos \chi  & 0 
\\
0 & 0 & 0}
\\
&
\quad
\times
\mtx{
 - i\cos^2 \chi 
 & -  e^{ i  (\theta  - \phi) }  \sin\chi  \cos \chi  & 0 
\\ 
 e^{-  i  (\theta  - \phi) }  \sin\chi  \cos \chi  &
   i\cos^2 \chi   & 0 
\\
0 & 0 & 0}
\der_\rho \chi
\der_\varphi \phi
\der_z\theta 
\\
&
= 
+ \sin \chi \cos \chi 
\der_\rho \chi
\der_\varphi \phi
\der_z\theta 
\mtx{ 1 && \\ & 1 & \\ &&0},
\end{split}
\end{equation}
and so on.
Therefore, we obtain
\begin{equation}
\begin{split}
\Gamma_5
    &
    =
\frac{ i N_{\rm C}}{48 \pi^{2}} \int d^{4} x \varepsilon^{i j k} \operatorname{Tr}\left[\left(L_{i} L_{j} L_{k}+R_{i} R_{j} R_{k}\right)  
\Omega_0 \right]  |_{\xi = 1} \\
&
=
- \frac{2 \cdot 3! \cdot N_{\rm C}}{48 \pi^{2}} \int d^{3} \bs{x} 
\sin \chi \cos \chi 
\der_\rho \chi
\der_\varphi \phi
\der_z\theta 
\int_0^{2 \pi } d t (-1) \,.
\end{split}
\end{equation}
The spatial integral is evaluated as
\begin{equation}
\begin{split}
&
 \int_0^\infty d\rho \int_0^{2\pi} d\varphi \int_{-\infty}^{\infty}dz 
 \sin \chi \cos \chi 
\der_\rho \chi
\der_\varphi \phi
\der_z\theta  
\\
&
= 
2\pi \nu  \left(\int_0^{R}  d\rho 
 \int_{-\infty}^{\infty}dz \der_z\theta  
+ 
\int_{R}^\infty d\rho 
 \int_{-\infty}^{\infty}dz \der_z\theta  
\right)
 \sin \chi \cos \chi 
\der_\rho \chi
\\
&
= 
2\pi \nu \int_0^{\fr{\pi}{2}}  d\chi
 \int_{-\infty}^{\infty}dz \der_z\theta  
 \sin \chi \cos \chi 
\\
&
= 
 2\pi^2 \nu .
\end{split} 
\label{eq:intgral_pancale}
\end{equation}
Therefore, we find the WZW term for the pancake solutions is given by
\begin{equation}
\begin{split}
\Gamma_5
&
    =
\frac{ i N_{\rm C}}{48 \pi^{2}} \int d^{4} x \varepsilon^{i j k} \operatorname{Tr}\left[\left(L_{i} L_{j} L_{k}+R_{i} R_{j} R_{k}\right)  
\Omega_0 \right]  |_{\xi = 1} \\
&
=
- \frac{2 \cdot 3! \cdot N_{\rm C}}{48 \pi^{2}}  (2 \pi^2 \nu )  
\cdot (-2 \pi) 
= N_{\rm C} \pi \nu .
\end{split}
\end{equation}
It implies that the pancake solitons are fermions for $N_{\rm C}=3$.

Note that for the domain wall Skyrmion, the second term in the second line of eq.~\eqref{eq:intgral_pancale} does not vanish, and one can evaluate it as
\begin{equation}
    \int_R^\infty d\rho \int_{-\infty}^\infty dz \partial_z\theta \partial_\rho \chi \sin\chi \cos\chi
    =-2\pi \int_{\pi/2}^0 d\chi \sin\chi \cos\chi
    =\pi.
\end{equation}
It implies that the WZW term for the domain wall Skyrmion is $2\pi$, and thereby they are bosons.

Since the baryon number 
can be similarly calculated 
as
\begin{equation}
\begin{split}
  N_\text{B} & =  \fr{1}{24 \pi^2} 
   \int d^3 \bs{x} 
\epsilon^{ijk}  \tr (L_i L_j L_k)
\\
&
= 
  -  \fr{1}{2 \pi^2} 
  \int d^3 \bs{x} 
    \sin \chi \cos \chi 
\der_\rho \chi
\der_\varphi \phi
\der_z\theta 
\\
&
= 
  -  \fr{1}{2 \pi^2} 
  \int_0^\infty d\rho \der_\rho \chi \sin \chi \cos \chi 
 \int_0^{2\pi} d\varphi\der_\varphi \phi
 \int_{- \infty}^\infty dz
 \der_z\theta 
 \\
&
= 
\begin{cases}
  -  2 \nu 
  \displaystyle{\int_0^{\pi/2}}
d \chi
 \sin \chi \cos \chi  = - \nu, 
 & \text{for pancake solitons [Figure 3 (c)]}
 \\
   + 2 \nu 
  \displaystyle{\int_{\pi/2}^0 }
d \chi
 \sin \chi \cos \chi =  - \nu ,
  &
  \text{for hole solitons [Figure 5 (b)]},
\end{cases}
\end{split}
\end{equation}
we can relate the WZW term and 
the baryon number as 
\begin{equation}
    \Gamma_5 = - \pi N_{\rm C} N_{\rm B}  .
\end{equation}

\bibliographystyle{jhep}
\bibliography{reference.bib}


\end{document}